\documentclass{SciPost}

\hypersetup{
    colorlinks,
    linkcolor={red!50!black},
    citecolor={blue!50!black},
    urlcolor={blue!80!black}
}

\usepackage[bitstream-charter]{mathdesign}
\DeclareSymbolFont{usualmathcal}{OMS}{cmsy}{m}{n}
\DeclareSymbolFontAlphabet{\mathcal}{usualmathcal}

\fancypagestyle{SPstyle}{
\fancyhf{}
\lhead{\colorbox{scipostblue}{\bf \color{white} ~SciPost Physics }}
\rhead{{\bf \color{scipostdeepblue} ~Submission }}

\fancyfoot[C]{\textbf{\thepage}}
}

\usepackage{amsmath,amsthm,verbatim,
            mathtools,enumitem,etoolbox,graphicx,physics,
            booktabs}

\usepackage{subfigure}
\usepackage{orcidlink}

\definecolor{romared}{RGB}{142,0,28}
\newcommand{\pd}{\partial}
\newcommand{\be}{\begin{equation}}
\newcommand{\ee}{\end{equation}}

\def\be{\begin{equation}}
\def\ee{\end{equation}}
\newcommand{\beq}{\begin{eqnarray}}
\newcommand{\eeq}{\end{eqnarray}}

\usepackage{makecell}
\usepackage{soul}

\newcommand{\calg}{\mathcal G}

\newcolumntype{Y}{>{\centering\arraybackslash}X}

\newcommand{\boundeq}{\mathrel{\hat =}}

\begin{document}

\pagestyle{SPstyle}

\begin{center}{\Large \textbf{\color{scipostdeepblue}{
Pseudospectral implementation of the Einstein-Maxwell system
}}}\end{center}

\begin{center}\textbf{
Jorge Expósito Patiño\orcidlink{0009-0007-5136-3126}\textsuperscript{1$\star$},
Hannes R. Rüter\orcidlink{0000-0002-3442-5360}\textsuperscript{1} and
David Hilditch\orcidlink{0000-0001-9960-5293}\textsuperscript{1}
}\end{center}

\begin{center}
{\bf 1} CENTRA, Departamento de F\'{\i}sica, Instituto Superior
  T\'ecnico -- IST, Universidade de Lisboa -- UL, Avenida Rovisco Pais
  1, 1049-001 Lisboa, Portugal
 \\[\baselineskip]
 $\star$ \href{mailto:jorge.exposito.patino@tecnico.ulisboa.pt}{\small jorge.exposito.patino@tecnico.ulisboa.pt}
\end{center}

\section*{\color{scipostdeepblue}{Abstract}}
\textbf{\boldmath{%
 Electromagnetism plays an important role in a variety of
  applications in gravity that we wish to investigate. To that end, in
  this work, we present an implementation of the Maxwell equations
  within the adaptive-mesh pseudospectral numerical relativity code
  \textsc{bamps}. We perform a thorough analysis of the evolution
  equations as a first order symmetric hyperbolic system of PDEs.
  This includes both the construction of the characteristic variables for use
  in our penalty boundary communication scheme, as well as radiation
  controlling, constraint preserving outer boundary conditions which,
  for the first time in a numerical context, are shown to be
  boundary-stable. After choosing a formulation of the Maxwell
  constraints that we may solve for initial data, we move on to
  show a suite of numerical tests. Our simulations, both within the
  Cowling approximation, and in full non-linear evolution, demonstrate
  rapid convergence of error with resolution, as well as consistency
  with known quasinormal decay rates on the Kerr background. Finally
  we evolve the electrovacuum equations of motion with
  strong data, a good representation of typical critical collapse
  runs.
}}

\vspace{\baselineskip}

\vspace{10pt}
\noindent\rule{\textwidth}{1pt}
\tableofcontents
\noindent\rule{\textwidth}{1pt}
\vspace{10pt}
\section{Introduction}
\label{section:introduction}

In this paper we initiate a series of numerical studies of the
dynamics of the Einstein-Maxwell system. The Einstein-Maxwell system
represents the interaction of electromagnetic waves and gravity. It is
a realistic and consistent model that could have relevant, unexplored
astrophysical applications in the strong-field regime. Moreover, we
expect the model to allow us to learn about the interaction of matter
and gravity in a simplified setting. Questions of interest include
those concerning the effect of strong electromagnetic fields in the
dynamics and relaxation of black holes, the interplay between
electromagnetic and gravitational waves, including possible black hole
formation from large-amplitude electromagnetic waves and, more
immediately, the study of critical collapse. Particularly in the
context of this last topic, where the numerical evolutions become
highly challenging, a clear and robust set of code verification and
validation tests is crucial. When tuned {\it very} close to the
threshold of collapse, different methods have occasionally been seen
to result in different physical results (see~\cite{Baumgarte:2019fai,
  Mendoza:2021nwq, Reid:2023jmr} for a detailed discussion). The
purpose of the present paper is therefore to present a suite of
verification and validation tests of our implementation so that we can
confidently claim validity of our results while anticipating potential
future difficulties.

Several implementations of the Maxwell equations have been employed in
the context of numerical relativity. Continuum formulations of the
equations have been built, for example, upon the work
of~\cite{Komissarov:2007wk} or~\cite{Alcubierre:2009ij}. Numerical
simulations have been performed to understand the effect of the
late-stages of inspiral and merger on the electromagnetic field with
uncharged compact objects~\cite{Palenzuela:2009yr}. A series of works
have been carried out on the effect of the electromagnetism on the
inspiral, merger and stability of charged black
holes~\cite{Zilhao:2012gp,Zilhao:2013nda,Zilhao:2014wqa}. Charged
black holes have also been treated in extreme
configurations~\cite{Bozzola:2022uqu,Bozzola:2023nzo,Smith:2024nwq,Smith:2024vka}
with a view to examining the effect of the Maxwell field on zoom-whirl
behavior and extremality.

Here we present an implementation of the Einstein-Maxwell system
inside the \textsc{bamps}
code~\cite{Hilditch:2015aba,Renkhoff:2023nfw}. The code uses the
PseudoSpectral Collocation (PSC) method which has been shown
empirically to result in highly accurate simulations when dealing with
smooth spacetimes~\cite{Grandclement:2007sb,Szilagyi:2014fna}. This is
expected on the basis of exponential convergence rates of the PSC, as
compared with power-law convergence provided by finite difference (FD)
approaches. To the best of our knowledge, this is the first PSC
implementation of the Einstein-Maxwell system (although
see~\cite{Kim:2024mau} for a Cowling implementation of force-free
electrodynamics). Besides the PSC method, we employ constraint damping
along with radiation controlling, constraint preserving boundary conditions,
both of which suppress unphysical behavior in our time evolutions. We prove
boundary stability for our choice of boundary conditions, which is the first
time ever that boundary stability has been shown for the Einstein-Maxwell
system.

The paper is organized as follows. In
section~\ref{section:continuum_analysis} we give a continuum analysis
of the evolution equations, including a discussion of adjustments to
the system to aid numerical evolution. We present the constraint
damping scheme, the characteristic variables of the system, the
formulation as an initial boundary value problem including our
boundary conditions, and our method for specifying constraint
satisfying initial data. In section~\ref{section:implementation} we
describe the implementation in detail. A brief description of the PSC
method is given, and the penalty and Bjørhus methods for the boundary
conditions are outlined, as well as our symmetry reduction (cartoon)
method. In section~\ref{section:numerical_results} we present various
tests that ensure correctness of distinct parts of the
implementation. We show both verification through convergence tests
and also validation by checking agreement with well established physics,
such as quasinormal modes around a Kerr black hole. Furthermore, we
include a simulation representative of the type of runs we will
perform in the study of the threshold of black hole formation. Our
conclusions are contained in
section~\ref{section:conclusions}. We use geometric units ($G=c=1$)
and Gaussian units for the Maxwell equations ($\varepsilon_0 =
1/4\pi$).

\section{Continuum analysis}
\label{section:continuum_analysis}

In this section we give a continuum formulation of the Maxwell
equations, compute the associated characteristic variables,
demonstrate boundary stability of the system with our boundary
conditions, and examine the Maxwell constraints in a form amenable to
numerical solution.

\subsection{Einstein-Maxwell system and constraint damping}

The equations we want to evolve are the Einstein-Maxwell system,
generally expressed as
\begin{gather}
  R_{\mu\nu} - \frac 12 R g_{\mu\nu} = 8\pi T_{\mu\nu} \, , \\
  \label{eq:maxwell_F}
  \nabla_\mu F^{\mu\nu} = -4\pi j^\nu, \\
  \label{eq:maxwell_F_hodge}
  \nabla_\mu (F^*)^{\mu\nu} = 0 \,, \\
  T_{\mu\nu} = \frac 1 {4\pi} \bigg(F_{\mu\lambda} F_{\nu}{}^{\lambda} 
    - \frac 1 4 g_{\mu\nu} F_{\rho\sigma} F^{\rho\sigma}\bigg) \, .
\end{gather}
$F_{\mu\nu}$ is the Faraday tensor,
$(F^*)^{\mu\nu} = -\frac12 \epsilon^{\mu\nu\rho\sigma} F_{\rho\sigma}$
is its Hodge dual and~$j^\mu$ is the charge four-current density. We use
the convention that $\epsilon^{0123} = -1/\sqrt{-g}$ and 
$\epsilon_{0123} = + \sqrt{-g}$ following \cite{Alcubierre:2009ij}.
The
gravity-electromagnetism interaction comes both from the
energy-momentum tensor for Einstein equations, which is quadratic on
the Faraday tensor, as well as from the covariant derivatives on the
Maxwell equations. For numerical relativity we want an initial value
formulation, which leads us to define the (foliation-dependent)
electric and magnetic fields
\begin{align}
    E^\mu = -n_\lambda F^{\lambda \mu}, \quad B^\mu = - n_\lambda (F^*)^{\lambda\mu} \, ,
\end{align}
with~$n^\mu$ the vector normal to constant time slices. In terms of
these, the evolution equations are~\cite{Alcubierre:2009ij}
\begin{align}
  \partial_t E^i &= \mathcal L_\beta E^i 
                   + {}^{(3)}\epsilon^{ijk}\partial_j(\alpha B_k) 
                   + \alpha K E^i - 4\pi \alpha {}^{(3)} j^i \, ,
                   \label{eq:evolution_E}\\
  \partial_t B^i &= \mathcal L_\beta B^i 
                   - {}^{(3)}\epsilon^{ijk}\partial_j(\alpha E_k) + \alpha K B^i \, ,
                   \label{eq:evolution_B}
\end{align}
where we introduced the spatial metric,
\begin{align}
\gamma_{\mu\nu} =  g_{\mu\nu} + n_\mu n_\nu \, ,
\end{align}
the lapse, $\alpha = -n_0$, the shift, $\beta^i = -\alpha n^i$, the
trace of the extrinsic curvature,
\begin{align}
K = -\frac 1 2 \gamma^{ij} \mathcal L_n \gamma_{ij} \, ,
\end{align}
the spatial charge-current density,
\begin{align}
{}^{(3)} j^i = \gamma^i{}_\mu j^\mu \, ,
\end{align}
and the timelike projection of the Levi-Civita tensor
\begin{align}
{}^{(3)}\epsilon_{\mu\nu\lambda} = n^\sigma \epsilon_{\sigma\mu\nu\lambda} \, .
\end{align}
As a reminder, the Lie derivative of a vector is
\begin{align}
  \mathcal L_\beta E^i = \beta^k \partial_k E^i - E^k \partial_k \beta^i\,.
\end{align}
and for a scalar, it reduces to the usual directional derivative,
\begin{align}
  \mathcal L_\beta \mathcal G_E = \beta^k \partial_k \mathcal G_E\,.
\end{align}

Observe that, in vacuum, equations \eqref{eq:evolution_E} and \eqref{eq:evolution_B}
have the following symmetry
\begin{align}
    E^i \to B^i, \quad B^i \to - E^i \, ,
\end{align}
which simplifies finding the magnetic equation once the electric one
is given. In any case, here we will always present both for
completeness. The Maxwell equations also provide two constraint
equations that have to be satisfied in all slices of the
foliation. Those are
\begin{align}
  \mathcal G_E := D_i E^i -4\pi \rho_E = 0, \quad
  \mathcal G_B := D_i B^i = 0 \, . \label{eq:em_constraints}
\end{align}
The term~$\rho$ is defined as~$\rho_E = -n_\mu j^\mu$ and $D_i$ is the Levi-Civita covariant
derivative of the spatial metric, defined as
\begin{align}
  D_i E^k = \gamma_i{}^\mu\gamma^k_l \nabla_\mu E^l\,,
\end{align}
for any spatial vector $E^k$.
If the evolution equations are exactly satisfied, these constraints
evolve as
\begin{align}
  \partial_t \mathcal G_E - \mathcal L_\beta \mathcal G_E 
  &=  \alpha K \mathcal G_E \, ,\\
  \partial_t \mathcal G_B - \mathcal L_\beta \mathcal G_B 
  &=  \alpha K \mathcal G_B \, .
\end{align}
Therefore, if the constraints are satisfied in the initial data, the
constraint violation remains zero in the continuum limit. Of course,
in numerical calculations a combination of truncation error and
round-off introduces constraint violations. We therefore modify the
equations in such a way that:
\begin{enumerate}
    \item The equations are the same when the constraints are zero.
    \item Constraint violations are suppressed in the evolution.
\end{enumerate}
Following this idea various different formulations of the Maxwell
equations have appeared in the literature as models for GR, see for
instance~\cite{Knapp:2002fm,Hilditch:2013sba}. The specific
modification we employ is analogous to the usual method for the
generalized harmonic gauge (GHG) or Z4 formulations of
GR~\cite{Brodbeck:1998az,Bona:2002fq,Gundlach:2005eh,Pretorius:2005gq},
divergence cleaning for MHD~\cite{Dedner02} and very similar to the
approach of~\cite{Reid:2023jmr}. We add two new terms to the Maxwell
equations, so that Eq.~\eqref{eq:maxwell_F} becomes
\begin{align}
    \nabla_\mu F^{\mu\nu} + \nabla^\nu Z_E - 2 \kappa n^\nu Z_E = -4\pi j^\nu \, .
\end{align}
This modification results in a new evolved variable $Z_E$ but has the 
advantage of better constraint suppressing properties.  $\kappa$ is a free
parameter that controls the amount of constraint damping, and in
section~\ref{subsec:results_flat} we compare evolutions with different
values of $\kappa$. 
We proceed analogously with Eq.~\eqref{eq:maxwell_F_hodge}, adding
two new terms with the equivalent variable $Z_B$ for the
magnetic field.
With these modifications the equations for the
constraints become slightly more complicated, but in Minkowski space
their evolution equations reduce to the simple form
\begin{align}
    \square \mathcal G_E + 2\kappa \partial_t \mathcal G_E = 0 \, , \quad
    \square \mathcal G_B + 2\kappa \partial_t \mathcal G_B = 0 \, ,
\end{align}
which is the standard equation for a damped wave.  We study the
constraint subsystem in generality in
section~\ref{subsec:continuum_analysis_characteristic}.

To summarize, our final formulation of the Maxwell evolution equations
is
\begin{align}
  \label{eq:maxwell_3_plus_1_E}
  \partial_t E^i &= \mathcal L_\beta E^i  -\alpha D^i Z_E
                   + {}^{(3)}\epsilon^{ijk}\partial_j(\alpha B_k)
                   +  \alpha K E^i - 4\pi \alpha {}^{(3)} j^i \, ,\\
  \label{eq:maxwell_3_plus_1_B}
  \partial_t B^i &= \mathcal L_\beta B^i -\alpha D^iZ_B- {}^{(3)}\epsilon^{ijk}\partial_j(\alpha E_k)
                   +\alpha K B^i \, ,\\
  \label{eq:maxwell_3_plus_1_ZE}
  \partial_t Z_E  &= \mathcal L_\beta Z_E -\alpha \left(D_i E^i - 4\pi \rho\right)
                    - 2\kappa \alpha Z_E\,,\\
  \label{eq:maxwell_3_plus_1_ZB}
  \partial_t Z_B  &= \mathcal L_\beta Z_B -\alpha D_iB^i - 2\kappa \alpha Z_B\,.
\end{align}

\subsection{Characteristic analysis of the Maxwell system}
\label{subsec:continuum_analysis_characteristic}

In this and the next subsection, we establish basic PDE properties of
our formulation of the Maxwell equations. We give only a very brief
overview of the underlying theory threaded with these calculations,
but a detailed exposition can be found in the
textbooks~\cite{kreiss1989initial,Kreiss:2006mi} or in the excellent
review article~\cite{Sarbach:2012pr}.

We can only hope for solutions of a numerical approximation scheme to
converge to solutions of the underlying PDEs if the continuum problem
is well-posed. Well-posedness is the requirement that, at least locally,
the PDE problem has unique solutions that depend continuously, in a
suitable sense, on the given data. The system
\begin{align}
  \partial_t{u}={A}^p(x,{u})
  \partial_p{u}+{S}(x,{u})\,,\label{eq:FT1S}
\end{align}
is said to be symmetric hyperbolic if there exists a symmetric,
positive definite matrix~${H}(x,{u})$, called a
symmetrizer, such that~${H}{A}^p$ is symmetric for
each~$p$. Symmetric hyperbolic systems are naturally associated with
an energy density~$\mathcal{E}={u}^T{H}{u}$,
whose existence is a necessary and sufficient condition for the
system to be symmetric hyperbolic. Subject to smoothness requirements
on initial data and given coefficients, systems of PDEs that are
symmetric hyperbolic have a well-posed initial value problem. We are
interested in the combined Einstein-Maxwell system. Since, however,
the two sets of equations are coupled only through non-principal
(lower derivative) terms, and in the hyperbolic context basic PDE
properties are determined by the principal part we may restrict our
attention to the Maxwell equations alone, assuming that the background
spacetime is sufficiently smooth. An equivalent analysis of the GHG
equations of motion as implemented in~\textsc{bamps} can be found
in~\cite{Hilditch:2015aba} (see~\cite{Rinne:2006vv} for analysis of
the GHG boundary conditions). Due to this, we need to consider presently
only linear systems.

We now show that our formulation of the Maxwell system is symmetric
hyperbolic. We do this by finding the characteristic variables, and
constructing an energy norm. Finally, we also present a characteristic
analysis of the constraint sub-system.

Symmetric hyperbolic PDEs are automatically strongly hyperbolic, which
implies that they have a complete set of characteristic variables for
each unit spatial~$s_i$. To compute the characteristic variables, we
seek the left-eigenvectors of the principal
symbol~${P}^s\equiv {A}^ps_p$. The principal symbol can thus be
arrived at by considering the evolution equations~\eqref{eq:FT1S}, and
discarding derivatives transverse to~$s_p$ (from this point on denoted
as \emph{transverse}) and all non-principal terms. Given such an
eigenvector~${l}_\lambda$ with associated eigenvalue~$\lambda$, the
associated characteristic variable is given by the simple dot
product~${l}_\lambda\cdot{u}$. Characteristic variables have a
physical interpretation as the combination of evolved variables whose
solutions are, in a certain approximation, simple traveling waves of
speed~$\lambda$.

To compute the characteristic variables when working with tensorial
fields whose principal part involves primarily the spacetime metric,
it is convenient to make a~$2+1$ decomposition against the
vector~$s^i$ by defining the projection operator
\begin{align}
  \label{eq:projection_operator_q}
  q_{ij} = \gamma_{ij} - s_i s_j \,.
\end{align}
Under this decomposition, we naturally adapt the representation of the
variables to the~$s_p$-direction, so that the spatial tensor
components associated with directions transverse to~$s_p$ coincide
with the components perpendicular to~$s^i$, which we denote by capital
Latin indices. An index~$s$ denotes contraction with $s^i$, for
example $E^s = E^i s_i$.  With that decomposition, the principal
symbol of the equations can be represented by
\begin{align}
  \partial_t E^s &\simeq \beta^s \partial_s E^s
                   - \alpha \partial_s Z_E \, , \\
  \partial_t B^s &\simeq \beta^s \partial_s B^s
                   - \alpha \partial_s Z_B \, , \\
  \partial_t Z_E &\simeq \beta^s \partial_s Z_E
                   - \alpha \partial_s E^s \, , \\
  \partial_t Z_B &\simeq \beta^s \partial_s Z_B
                   - \alpha \partial_s B^s \, , \\
  \partial_t E^A &\simeq \beta^s \partial_s E^A
                   - \alpha {}^{(2)} \epsilon^A{}_B \partial_s B^B \, ,\\
  \partial_t B^A &\simeq \beta^s \partial_s B^A
                   + \alpha {}^{(2)} \epsilon^A{}_B \partial_s E^B \, .
\end{align}
The symbol~$\simeq$ here denotes ``equality up to non-principal terms
and transverse derivatives''. The 2d Levi-Civita tensor is defined as
\begin{align}
  {}^{(2)}\epsilon_{\mu\nu} := s^\lambda {}^{(3)} \epsilon_{\lambda \mu\nu}
  = n^\sigma s^\lambda \epsilon_{\sigma\lambda\mu\nu} \, .
\end{align}
The vectors with capital Latin indices are the transverse components to the
vector~$s^i$, i.e.
\begin{align}
 E^A = q^A{}_i E^i\,.
\end{align}

We have four scalar characteristic variables (the longitudinal ones)
and two two-vectors (the transversal ones) that are given by
\begin{align}
    \label{eq:transchars_longE}
    u_\pm^{\text{long}, E} &= \frac 1 {\sqrt 2} \left(\pm E^s + Z_E\right) \, , \\
    \label{eq:transchars_longB}
    u_\pm^{\text{long}, B} &= \frac 1 {\sqrt 2} \left(\pm B^s + Z_B\right) \, , \\
    u_\pm^A &= \frac 1 {\sqrt 2} \left( \pm E^A + {}^{(2)}\epsilon^A{}_C B^C\right) \, ,
    \label{eq:transchars_longA}
\end{align}
with speeds~$\lambda_\pm = - \beta^s \pm \alpha$, all agreeing with
the speed of light in the~$\pm s_p$-directions. Of these, the
transversal variables agree with those presented
in~\cite{Alcubierre:2009ij}, but because we work with a different
formulation, the longitudinal variables and speeds differ. The inverse
transformation is given by
\begin{align}
  E^i &= \frac 1 {\sqrt 2} \left[s^i \left(u_+^{\text{long}, E} - u_-^{\text{long}, E}\right)
        + q^i{}_A (u^A_+ - u^A_-)\right] \, , \\
  B^i &= \frac 1 {\sqrt 2} \left[\left(u_+^{\text{long}, B} - u_-^{\text{long}, B}\right)
        - {}^{(2)}\epsilon^i{}_A \left(u^A_+ + u^A_-\right)\right] \, , \\
  Z_E &= \frac{1}{\sqrt 2}\left(u_+^{\text{long}, E}+  u_-^{\text{long}, E}\right) \, , \\
  Z_B &= \frac{1}{\sqrt 2}\left(u_+^{\text{long}, B}+  u_-^{\text{long}, B}\right) \, .
\end{align}
The characteristic variables form a complete basis and have real
characteristic speeds, which shows that the system is strongly
hyperbolic, we can build an energy norm independent of $s^i$, showing
that the system is also symmetric hyperbolic.
\begin{align}
    \mathcal E :={}& \left(u_-^{\text{long}, E}\right)^2 +
    \left(u_+^{\text{long}, E}\right)^2+\left(u_-^{\text{long}, B}\right)^2 +\left(u_+^{\text{long}, B}\right)^2
    +  u_+^Au_{+A} + u_-^Au_{-A}
    \nonumber\\
     ={}&  E^kE_k + B^k B_k + (Z_E)^2 + (Z_B)^2 \, .
\end{align}

Observe that this differs from the physical (ADM) energy density by
constraint addition, but that it is this quantity that allows us to
construct a norm in solution space as
\begin{align}
    \norm{u} =  \int_\Omega \mathcal{E} (u) \epsilon\,.
\end{align}
The integral is calculated over a spatial slice of constant~$t$ with the
appropriate induced volume form.

Besides their direct physical interpretation, the characteristic
variables play an important role in the analysis of the initial
boundary value problem (IBVP) of symmetric hyperbolic systems. At the
boundary, each characteristic variable may be classified as either in-
or outgoing, depending on the sign of the associated speed. In
particular, to obtain boundary conditions that do not introduce
constraint violation in the domain, we also study the characteristics
of the constraint sub-system.

It turns out~(see~\cite{Reula:2004xd}, or~\cite{Hilditch:2013ila} for
a version of the result for constrained Hamiltonian systems) that a
necessary condition for strong-hyperbolicity of the full system is
that the constraint subsystem itself be strongly-hyperbolic. From the
results above, we know that we can find a complete set of
characteristic variables for the constraint subsystem. In our case,
with the constraint propagating and damping adjustments, the full
constraint subsystem is
\begin{align}
  \partial_t Z_E - \mathcal L_\beta Z_E =& -\alpha \mathcal G_E
    - 2\kappa \alpha Z_E \, ,\\
  \partial_t \mathcal G_E - \mathcal L_\beta \mathcal G_E =&
    - \alpha D^2 Z_E -  (D^i Z_E) (D_i \alpha ) + \alpha K \mathcal G_E \, , 
    \\
  \partial_t Z_B - \mathcal L_\beta Z_B =& -\alpha \mathcal G_B
    - 2\kappa \alpha Z_B \, , \\
  \partial_t \mathcal G_B - \mathcal L_\beta \mathcal G_B =&
    - \alpha D^2 Z_B -  (D^i Z_B) (D_i \alpha ) + \alpha K \mathcal G_B \,.
\end{align}
In order to find the characteristic variables, we introduce the
(formal) first order reduction with the following definitions
\begin{align}
    \mathcal C_i := \partial_i Z_E \, , \quad \bar {\mathcal C}_i :=
    \partial_i Z_B \, ,
\end{align}
which leads to two decoupled principal parts of the constraint
subsystem, given by
\begin{align}
  \partial_t Z_E
  &\simeq 0 \, , \\
  \partial_t \mathcal C^A
  &\simeq 0 \, ,\\
  \partial_t \mathcal G_E
  &\simeq \beta^s \partial_s \mathcal G_E
    - \alpha \partial_s \mathcal C^s \, , \\
  \partial_t \mathcal C^s 
  &\simeq \beta^s \partial_s \mathcal C^s 
    - \alpha \partial_s \mathcal G_E \,,
\end{align}
and
\begin{align}
  \partial_t Z_B 
  &\simeq 0 \, , \\
  \partial_t \bar{\mathcal C}^A 
  &\simeq 0 \, ,\\
  \partial_t \mathcal G_B 
  &\simeq \beta^s \partial_s 
  \mathcal G_B - \alpha \partial_s \bar{\mathcal C}^s \, , \\
  \partial_t \bar{\mathcal C}^s 
  &\simeq \beta^s \partial_s \bar{\mathcal C}^s 
  - \alpha \partial_s \mathcal G_B \, .
\end{align}
The variables~$Z_E$, $Z_B$, $\mathcal C^A$ and $\bar{\mathcal C}^A$
are clearly seen to be characteristic variables with zero speed. The
rest are
\begin{align}
  c^E_{\pm} ={}& \frac{1}{\sqrt{2}} \left(\pm \mathcal G_E + \partial_s Z_E\right) \,, \\
  c^B_{\pm} ={}& \frac{1}{\sqrt{2}} \left(\pm \mathcal G_B + \partial_s Z_B\right)\,,
\end{align}
with speeds~$\lambda_{\pm}= - \beta^s \pm \alpha$. The ones with
negative speed represent incoming constraint violation at the
boundary. We examine constraint preserving boundary conditions in more
detail in the next section.

\subsection{Outer boundary conditions}
\label{section:boundary_conditions}

Our numerical method uses a finite domain with a time-like boundary,
meaning we do not solve the Einstein-Maxwell system as an initial
value problem, but rather as an initial boundary value problem. In
order to obtain a well-posed IBVP, we have to choose appropriate
boundary conditions. Our guiding principle for the choice of boundary
conditions is to emulate as much as possible the behavior of the
initial value problem.

Given that the system is symmetric hyperbolic, an obvious choice for
boundary conditions that renders the IBVP well-posed would be
maximally dissipative boundary conditions~\cite{Hilditch:2013sba}. The
multidomain numerical approach we employ (see~\cite{Hes00,HesGotGot07}
for details) uses maximally dissipative boundary conditions to carve
up the computational domain into many small patches in which the
incoming data of each patch corresponds to the outgoing data from its
neighbors (see section~\ref{section:implementation}). Unfortunately, at
the outer boundary, we cannot simply employ maximally dissipative
boundary conditions. The reason for this is that generic boundary
conditions will not be compatible with the constraints of the
theory. We thus need to consider boundary conditions in that relate
properly to constraint violations.

The fact that the boundary is time-like is equivalent to the
algebraic condition at the boundary
\begin{equation}
 |\beta^s| < \alpha\,,
\end{equation}
where, here and throughout this section,~$s^i$ is the spatial normal to
the boundary. The incoming variables upon which boundary conditions can be freely imposed are
\begin{align}
    u_-^A\,,\quad u_-^{\text{long}, E}\,,\quad u_-^{\text{long}, B}\,,
\end{align}
and we impose two types of boundary conditions: those that specify
incoming physical data and those that ensure constraint preservation.

For the first type, we construct the Newman-Penrose scalar associated
with incoming electromagnetic radiation
\begin{align}
    \phi_0 = F_{\mu\nu} l^\mu m^\nu  = u_-^A m_A\, .
\end{align}
Up to non-principal terms, these characteristics do not affect the
incoming constraint violation, so we are free to choose maximally
dissipative boundary conditions. In particular we will fix the
incoming radiation to zero, which, in terms of the physical
$E^i$ and $B^i$ fields, is expressed as
\begin{align}
  \phi_0 \boundeq  0 \Rightarrow u_-^A \boundeq 0
  \Rightarrow -E^A + {}^{(2)} \epsilon^A{}_C B^C \boundeq 0\,,
  \label{eq:maxdiss}
\end{align}
where the symbol~``$\boundeq$'' means equality at the boundary. This
choice results in a difference with the initial value problem, since
back-scattering from waves outside the domain is neglected. The
difference, however, decreases as the outer boundary is placed farther in the
asymptotically flat zone. Alternatively, we could have chosen
Sommerfeld-type boundary conditions
\begin{equation}
 (\partial_t + \partial_s) \phi_0 \boundeq 0\,,
\end{equation}
which also results in a well-posed problem, or perhaps even higher
order derivative
generalizations~\cite{Buchman:2006xf,Sarbach:2007hd,Rinne:2008vn}
thereof. This would help to reduce reflections at the boundary. We
have chosen to go with condition~\eqref{eq:maxdiss} since the
interpretation in terms of incoming given data is more direct.

On the other hand, we have to impose boundary conditions such that
there is no incoming constraint violation through the boundary. Having
studied the constraint subsystem, we know that the incoming constraint
violation through a timelike boundary is given by the
variables~$c_-^E$ and~$c_-^B$, and we want to set them to zero on the
boundary. This is equivalent to the following Robin condition on the
fields
\begin{align}
  D_i E^i - \partial_s Z_E \boundeq 0\,,\quad
  D_i B^i - \partial_s Z_B \boundeq 0\,.
\end{align}
Up to non-principal terms, this is equivalent to Sommerfeld boundary
conditions on~$Z_E$ and~$Z_B$. These are {\it constraint preserving},
since starting with initial data that satisfies the constraints, at
the continuum level the solution of the IBVP will satisfy the
constraints too. Furthermore, they are {\it constraint absorbing},
since (with at most this many derivatives in the conditions) they help
minimize reflections of the constraints at the boundary 
(see~\cite{Sarbach:2007hd} for further discussion).

Working with those boundary conditions and our formulation of the
Maxwell equations, we show in this section that the system is boundary
stable. To the best of our knowledge this is the first such proof for
this specific setup. Since the system is also symmetric hyperbolic,
this in turn implies that the IBVP is well-posed in a generalized
sense~\cite{Kre70,Agr72,Met00}.

In a simulation, we will generally have a certain amount of numerical
error, which is characterized both by being small in comparison with
the solution (at least when it first appears) and very high frequency,
since the error at neighboring grid points is largely independent. If
we want to make sure our implementation is valid, we have to make sure
that the IBVP is “robust”, in the sense that small high-frequency
perturbations do not cause undesired effects, especially blow-up of
the solution. This is where the concept of boundary stability comes
in.

In order to study the behavior of the system under small perturbations
on the boundary, we allow in our analysis for small arbitrary given
data in the boundary
\begin{gather}
 \phi_0 \boundeq g_1\,, \quad c^E_- \boundeq g_2\,, \quad c^B_-\boundeq g_3\,.
\end{gather}
Roughly, a system is boundary stable if the incoming Fourier-Laplace
modes are bounded by the given data at the boundary
\cite{kreiss1989initial}. The precise definition is given
in~\eqref{eq:bound_stability}, once the necessary terms have been
defined. The Fourier-Laplace transformation is used to transform the
PDE system into an ODE system with initial data at the boundary. In
the following we study how this applies to the Einstein-Maxwell case.

We want to study small, high-frequency perturbations, which allows to
make a couple of simplifying approximations. Since the perturbations
are small, we can study the linear system around any given
solution. The Maxwell part of the equations is already linear by
itself, so this is the same as fixing the background metric. On the
other hand, since they are very high-frequency, we can consider the
background terms constant, and neglect the forcing terms, i.e. the
ones that do not contain derivatives. This is the case since for any
high-frequency mode $u \sim e^{k^\mu x_\mu}$
\begin{align}
  \partial_\mu u \sim k_\mu u
  \Rightarrow \sqrt{\delta^{\mu\nu}(\partial_\mu u)(\partial_\nu u)} \gg u\,.
\end{align}
Under those
simplifications, the system of evolution 
equations~\eqref{eq:maxwell_3_plus_1_E}-\eqref{eq:maxwell_3_plus_1_ZB} 
is given by
\begin{align}
    \partial_t E^i &= \beta^k \partial_k E^i 
    - \alpha D^i Z_E + \alpha\,
    {}^{(3)}\epsilon^{ijk}\partial_j B_k\,, \\
    \partial_t B^i &= \beta^k \partial_k B^i- \alpha D^i Z_B 
    -\alpha\, {}^{(3)}\epsilon^{ijk}\partial_j E_k\,, \\
    \partial_t Z_E &= \beta^k \partial_k Z_E
    - \alpha D_i E^i\,, \\
    \partial_t Z_B &= \beta^k \partial_k Z_B- \alpha D_i B^i\,.
\end{align}
For a given point in the boundary, we can choose coordinates such that
the frozen metric has the form~\cite{Ruiz:2007hg}
\begin{align}
  \mathring g_{\mu\nu}dx^\mu dx^\nu = -dt^2 + (dx + \beta dt)^2 + dy^2 + dz^2\,,
\end{align}
and~$x=0$ corresponds to the boundary, with $\partial_x$ being
incoming into the domain there. Furthermore, under the assumption that
the boundary is timelike, we have the condition
\begin{align}
    |\beta| < 1\,.
\end{align}
Generally, the form of our system is
\begin{align}
 \partial_t u = A^p \partial_p u\,, \quad B^p\partial_p u + Cu \boundeq Lg\,,
 \label{eq:IBVPcc}
\end{align}
the first equation representing the evolution, and the second one the
boundary conditions.  Since boundary conditions are only imposed on the
incoming characteristics, which in our case represent half of the
degrees of freedom, $B^p$ and $C$ are $4\times8$ matrices, and $g$ has
4 components. Observe that we allow a general linear
operator~$L$ on the given data. In boundary conditions that contain up
to one derivative of the solution, we allow up to one derivative,
within the boundary, of the given data within this operator. This
affects the specific form of the norms that are obtained within the
well-posedness result (see~\cite{Ruiz:2007hg} for a detailed
explanation.)

By performing a Laplace-transform in time and a
Fourier-transform in the space tangential to the boundary
\begin{align}
  \hat f(x, s, \omega_y, \omega_z)
  := \int f(t, x, y, z) e^{-i\omega_y y - i\omega_z z - st}\; dt dy dz\,,
\end{align}
we can transform~\eqref{eq:IBVPcc} into an ODE system with initial data at the
boundary.

The transformation of system~\eqref{eq:IBVPcc} is
\begin{align}
  \partial_x \hat u = M \hat u\,, \quad
  N_1 \partial_x \hat u + N_2 \hat u
  \boundeq \hat L\hat g\,.\label{eq:FLsystem}
\end{align}
For the Einstein-Maxwell system, with an appropriate ordering of the
variables in the solution vector, given by
\begin{align}
  \hat u = \left[\begin{array}{c}\hat E^x \\ \hat B^y \\ \hat E^z \\
                   \hat Z_E \\ \hat B^x \\ \hat E^y \\ \hat B^z \\
                   \hat Z_B\end{array}\right]=
                   \left[\begin{array}{c}\hat v_E \\
                   \hat v_B\end{array}\right]\,,
\end{align}
with $\hat v_E$ and $\hat v_B$ vectors with four components,
we can split the whole system into two identical copies.
In these variables,
equations~\eqref{eq:FLsystem} are of the form
\begin{gather}
  \partial_x\left[\begin{array}{c}\hat v_E \\ \hat v_B\end{array}\right] =
  \left[\begin{array}{cc}\bar M & 0 \\ 0 & \bar M\end{array}\right] \left[\begin{array}{c}\hat v_E \\
  \hat v_B\end{array}\right]\,,\\
     \left[
       \begin{array}{cc} \bar N_1 & 0 \\
                                0 & \bar N_1\end{array}
     \right] 
     \partial_x \left[\begin{array}{c}\hat v_E \\ \hat v_B\end{array}\right] 
     + \left[\begin{array}{cc}
          \bar N_2 & 0 \\ 
                 0 & \bar N_2\end{array}\right] 
     \left[\begin{array}{c}\hat v_E \\ 
                           \hat v_B\end{array}\right]
     \boundeq
     \left[\begin{array}{cc}
        \bar L & 0 \\ 
             0 & \bar L\end{array}\right] 
     \left[\begin{array}{c}\hat g_E \\ \hat g_B\end{array}\right]\,.
\end{gather}
This allows us to study just one half of the system. Showing boundary
stability of one of the two copies will automatically demonstrate
boundary stability of the whole system.
The matrices given above are
\begin{align}
    \bar M = \frac{1}{1-\beta^2}
    \left[\begin{array}{cccc}
     -s \beta & -i \beta  \omega  & -i \omega  & -s \\
     -i \beta  \omega  & -s \beta & s & i \omega  \\
     i \omega  & s & -s \beta & -i \beta  \omega  \\
     -s & -i \omega  & -i \beta  \omega  & -s \beta \\
    \end{array}\right]\,.
\end{align}
and
\begin{gather}
  \bar N_1 = \left[\begin{array}{cccc} 0 & 0 & 0 & 0 \\
                   -1 & 0 & 0 & -1
                  \end{array}\right]\,, \\
  \bar N_2  =
                \left[\begin{array}{cccc} 0 & -1 & 1 & 0 \\
                   0 & 0 & -i \omega & 0
                  \end{array}\right]\,,\\
  \bar L = \left[\begin{array}{cc} 1 & 0 \\ 0 & s \end{array}\right]\,.
\end{gather}
The boundary conditions can be transformed into a purely algebraic
equation by substituting the equation for the $x$ derivatives
\begin{align}
 &\bar N_1 \bar M \hat v_E  + \bar N_2 \hat v_E \boundeq \bar L \hat g_E \nonumber\\ &\Rightarrow
 \bar{L}^{-1}(\bar N_1 \bar M + \bar N_2) \hat v_E = \bar X\hat v_E  \boundeq \hat g_E\,.
\end{align}
which for the Einstein-Maxwell case results in
\begin{align}
 \label{eq:Xbar}
  \bar X = \left[\begin{array}{cccc} 0 & -1 & 1 & 0\\
          \frac{1}{1-\beta} & \frac{i\omega}{s(1-\beta)} 
                            & \frac{i\omega\beta}{s(1-\beta)} & \frac{1}{1-\beta}
                \end{array}\right]\,.
\end{align}
As we can see, this matrix is not square, and in particular it is not
injective, so we cannot find the full solution at the boundary solely
given by the boundary data. This is not an accident of the
Einstein-Maxwell case. Indeed, it will happen in general, since the boundary
conditions only give data for incoming characteristics
variables. Physically, only half of the degrees of freedom are
incoming and are specified by the given data at this boundary.

We can restrict our analysis to the variables that do depend on the
given data by looking at the eigenvectors of~$\bar M$ with negative
real part eigenvalues. In general, the solutions of
\begin{align}
 \partial_x \hat v_E = \bar M\hat v_E\,,\nonumber
\end{align}
are of the form
\begin{align}
 \hat v_E = R^{(-)} \sigma_-(x, s,\omega) + R^{(+)} \sigma_+(x, s,\omega)\,,
\end{align}
where~$R^{(+)}$ is the matrix of eigenvectors whose eigenvalues have
positive real part, $R^{(-)}$ consists of those whose eigenvalues have
negative real part and $\sigma_\pm$ are some vectors.  Generally we
could have eigenvectors with zero real part, in which case we would
say that the boundary is characteristic with respect to some of the
variables, but in our case the assumption of time-like boundary
ensures that will not be the case.  By restricting to the subspace of
negative real part eigenvalues, the matrix~$X$ becomes injective.  We
can find the relation of the solution to the given data at the
boundary by
\begin{align}
 \hat v_E \boundeq R^{(-)} \left(X R^{(-)} \right)^{-1} \hat g_E =: F \hat g_E\,.
\end{align}
The condition of boundary stability is that there exists a bound
\begin{align}
 |\hat v_E(0, s,\omega)| < \delta |\hat g_E(s, \omega)|\,,
 \label{eq:bound_stability}
\end{align}
for some~$\delta > 0$. In this case, showing boundary stability is
equivalent to showing that the matrix~$F$ relating the physical and
given data is bounded.

In the Einstein-Maxwell case, the matrix~$M$ has two distinct
eigenvalues, given by
\begin{align}
    \tau_\pm = - \gamma^2(s\beta \mp \lambda)\,,
\end{align}
where we have defined
\begin{align}
  \gamma := (1- \beta^2)^{-1/2}\,,\quad
  \lambda := \sqrt{s^2 + \gamma^{-2}\omega^2}\,,
\end{align}
These eigenvalues appear also in the general relativity case, and it
has been shown~\cite{Hilditch:2016xos} that
\begin{align}
  \text{Re}(\tau_-) < 0 < \text{Re}(\tau_+)\,,\quad
  \text{given}\quad \text{Re}(s) \geq 0  ,  \; |\beta| < 1\,,
\end{align}
therefore the incoming variables are the ones associated
with~$\tau_-$. The incoming part of the solution is
\begin{align}
    \hat v_E = \underbrace{\left[\begin{array}{cc}
        \beta\omega^2 + s\lambda  & -i\omega(s\beta - \lambda) \\
        i\omega(s\beta - \lambda) & -(\beta\omega^2 + s\lambda)\\
        0                         & s^2 + \omega^2\\
        s^2 + \omega^2            & 0
                               \end{array}\right]}_{R^{(-)}}
        \left[\begin{array}{c}\sigma_1 \\
        \sigma_2\end{array}\right] e^{\tau_-x}\,.
\end{align}
We want to show a bound of~$\hat v_E$, and for that we show that the
absolute value of all matrix elements are bounded. To do so it is
useful to introduce the following variables
\begin{gather}
  k = \sqrt{|s|^2 + \omega^2}\,,\\ s' = s/k\,,\\
  \omega' = \omega/k\,,\\
  \lambda' = \lambda/k\,,\\ \tau' = - \tau_-/k\,,
\end{gather}
in such a way that all the primed variables have bounded absolute
value. With those variables we can rewrite the matrix relating given
data to physical fields at the boundary as
\begin{align}
   F = -\frac{1}{(s'+\lambda')^2}
   \left[
    \begin{array}{cc}
    \frac{i \omega ' \left(\lambda '+s'\right)}{(\beta +1) \gamma ^2} &
   \frac{s' \left(\lambda '+s'\right)}{(\beta +1) \gamma ^2} \\
    \frac{\left(\lambda '+s'\right) \left(\beta  \lambda '+s'\right)}{(\beta
   -1) (\beta +1)^2 \gamma ^2} & \frac{i s' \omega '}{(\beta +1)^2 \gamma
   ^4} \\
    -\frac{\left(\lambda '+s'\right) \left(\lambda '+\beta  s'\right)}{(\beta
   -1) (\beta +1)^2 \gamma ^2} & \frac{i s' \omega '}{(\beta +1)^2 \gamma
   ^4} \\
 0 & \frac{s' \left(\lambda '+s'\right)}{(\beta +1) \gamma ^2} \\
    \end{array}
    \right]
    \,.
    \label{eq:bounded_matrix_simple}
\end{align}
It is well known~\cite{Kreiss:2006mi, Rinne:2006vv, Ruiz:2007hg} that
the combination~$s' + \lambda'$ is bounded below by a positive value,
but for the sake of completeness, we show it here.

If~$\omega = 0$, then
\begin{align}
 |s' + \lambda'| = \frac{|s + |s||}{|s|} = |e^{i\arg{s}} + 1| =
 \cos \left(\frac{\arg s} 2\right) \geq \sqrt2\,, \quad \text{for} \quad \text{Re}(s) \geq 0\,.
\end{align}
On the other hand, if $\omega \neq 0$ we introduce $\zeta = s/\omega$
\begin{align}
 s' + \lambda' = \frac{\zeta + \sqrt{\zeta^2 + \gamma^{-2}}}{\sqrt{|\zeta|^2 + 1}}\,.
\end{align}
Assuming there is no such bound, there must exist a sequence
$\zeta_n\to \zeta_*$ in the half complex plane Re$(\zeta)\geq 0$ such
that
\begin{align}
  \frac{\zeta_n + \sqrt{\zeta_n^2 + \gamma^{-2}}}{\sqrt{|\zeta_n|^2 + 1}}
  \to 0\Rightarrow \zeta_* + \sqrt{\zeta_*^2 + \gamma^{-2}} = 0
  \Rightarrow \zeta_*^2 = \zeta_*^2 +\gamma^{-2} \Rightarrow \gamma^{-2} = 0\,,
\end{align}
and since~$\gamma^{-2}\neq 0 $ when the boundary is timelike
($|\beta| \leq 1$), we reach a contradiction, thus a positive lower
bound must exist.

Since the matrix entries are expressed in terms of primed variables,
which are bounded, and~$|s' + \lambda'|$ is bounded below by a
positive value, the matrix is bounded element-wise. Therefore, the
solution at the boundary is bounded by the given data, which in this half of
the system is
\begin{align}
    |\hat v_E(0, s, \omega)| \leq \delta |\hat g_E(s, \omega)|\,,
\end{align}
for some~$\delta > 0$. Since the two halves of the system are the same, this
bound applies to the whole system, and we have boundary stability.  As
mentioned before, since the system is symmetric hyperbolic, the
theorems of~\cite{Kre70,Agr72,Met00} ensure with sufficient
smoothness that the system is well-posed in the generalized sense.

\subsection{Initial Data}

In order to run free-evolution simulations we need constraint
satisfying initial data, meaning that we have to solve the constraint
equations~\eqref{eq:em_constraints} in a curved background at the
same as solving the gravity constraints. The full system of
constraints is
\begin{align}
    \mathcal H &= {}^{(3)}R + K^2 - K_{ij} K^{ij} - 16\pi\rho = 0 \, ,\\
    \mathcal M^i &= D_j\left(K^{ij} - \gamma^{ij} K\right) - 8\pi S^i = 0 \, ,\\
    \mathcal G_E &= D_i E^i - 4\pi\rho_E = 0 \, ,\\
    \mathcal G_B &= D_i B^i = 0 \, ,
\end{align}
where
\begin{align}
  \rho = \frac1{8\pi} \left(E^2 + B^2\right), \\
  S_i = \frac1 {4\pi} {}^{(3)}\epsilon_{ijk} E^j B^k \, .
\end{align}
are the contractions of the energy momentum tensor that appear in
a~$3+1$ decomposition.

The constraints constitute an underdetermined system of equations, so
we have to make choices for some of the variables and then solve the
constraint equations for the rest. We choose to do so with the
decomposition of the electromagnetic fields given by
\begin{align}
    E_i = \psi^{-2}\left(\tilde E_i - \partial_i\varphi_E\right) \, , \\
    B_i = \psi^{-2}\left(\tilde B_i - \partial_i \varphi_B\right) \, ,
\end{align}
where~$\psi$ is the conformal factor to be found by solving the Hamiltonian
and momentum constraints, $\tilde E^i$ and~$\tilde B^i$
are freely specifiable variables, and~$\varphi_E$, $\varphi_B$ are the
variables to solve for. That decomposition reduces the constraints to
Poisson-type equations on the conformal geometry
\begin{align}
  \label{eq:maxwell_constraints_conformal_E}
  \bar D^2 \varphi_E ={}& \bar D_k \tilde E^k - 4\pi \psi^6\rho_E \, , \\
  \label{eq:maxwell_constraints_conformal_B}
  \bar D^2\varphi_B ={}& \bar D_k \tilde B^k \, .
\end{align}
A conformal decomposition was already used in~\cite{Alcubierre:2009ij,
  Mendoza:2021nwq, Baumgarte:2019fai, Reid:2023jmr}. This constraint
formulation for the electromagnetic fields has a couple of advantages
when paired with a conformal decomposition of the gravity variables,
particularly in the electrovacuum case. Since in electrovacuum the
constraints~\eqref{eq:maxwell_constraints_conformal_E}
and~\eqref{eq:maxwell_constraints_conformal_B} depend only on
the conformal geometry, they can be solved independently, and in many
cases (in particular under the assumption of conformal flatness)
analytical solutions already exist. Presently we use analytical
solutions of the conformal constraints. These solutions have the added
advantage of an easy physical interpretation as non-linear extensions
of the corresponding linear solutions. This extension from linear
  to non-linear setting is non-unique, but we nevertheless refer to
electromagnetic data that is conformally a linear quadrupole as a
non-linear quadrupole, for example. On the gravity side, we use the
CTS approach~\cite{Baumgarte}.

\section{Implementation}
\label{section:implementation}
\begin{figure*}
    \centering
    \includegraphics[width=0.99\linewidth]{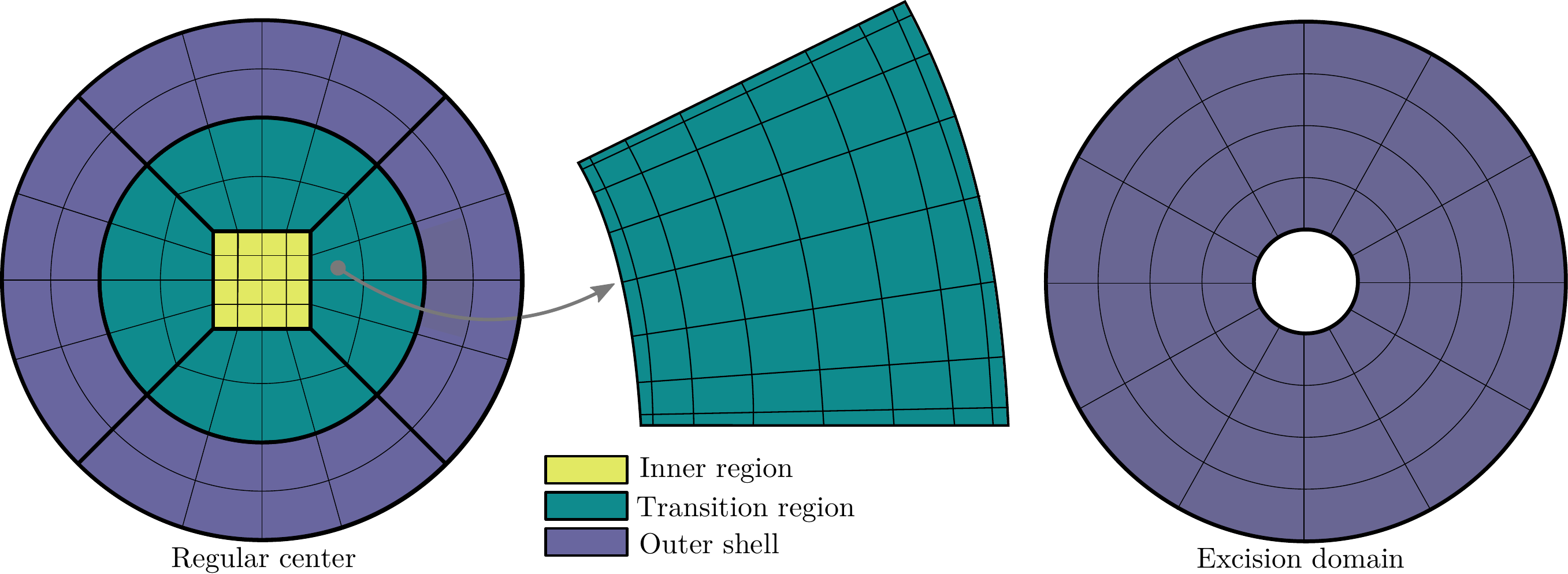}
    \caption{\label{fig:multipatch}In \textsc{bamps}, the domain is
      divided into three subdomains of different geometry. Each one of
      them is divided in patches. Inside each patch, the solution is
      represented as a linear combination of Chebyshev
      polynomials. The number of patches in the inner region is given
      to match the angular resolution outside.}
\end{figure*}

The Maxwell equations are implemented within the existing numerical
relativity code \textsc{bamps}, and they share the code structure and
tools with the existing code base. Below we give a more detailed
  discussion of details that relate directly to the Maxwell
  implementation but, for completeness, we begin with an overview of the
  general code structure. \textsc{bamps} is a 3d MPI parallel
  multidomain (spectral element) adaptive pseudospectral code taylored
  to the solution of first order symmetric hyperbolic PDEs.
  We employ the method of lines for time evolution, and
  the computational domain is divided into small
  patches, called spectral elements, on which an initial boundary
  value problem is solved. These elements can then be distributed to
  different computational cores, provided that appropriate data from
  neighboring elements are communicated. Since such data have to be
  provided on domains of codimension 1 within the computational
  domain, this approach is expected to scale well, which has been
  confirmed in earlier work, where
  the code has been shown to scale up to at least several
  thousand computational cores. That said, the experiments we have
  performed for this paper, in which we are primarily focused on
  establishing the validity of our Maxwell implementation, we used at
  most 768 cores. Data are communicated via a penalty
  method on characteristic variables. Various different options are
  possible at the outer boundary of the domain. When evolving black
  hole spacetimes we use black hole excision. The idea is that since
  the black hole boundary is pure outflow from the point of view of
  the PDEs, no continuum boundary conditions are required. We
  therefore simply monitor the boundary to be sure that it indeed
  remains outflow and apply no boundary conditions. To treat situations
  with symmetry we employ the cartoon method, in which a Killing vector
  associated with the symmetry is used reduce the dimensionality of
  the problem. Our mesh-refinement driver supports both~$h$ and~$p$
  refinement. For $h$~refinment, the number of spectral elements is
  increased, whereas for $p$~refinement, the number of grid-points per spectral
  element is increased. For a full description of the code,
  see~\cite{Hilditch:2015aba, Renkhoff:2023nfw}.

The discretization method is a multi-patch PseudoSpectral Collocation
(PSC) method. A schematic of a typical domain is presented in
Fig.~\ref{fig:multipatch}. Here we give a very abridged presentation
of the PSC method (for more detail see~\cite{fornberg1998practical,
  boyd2000}.)  Within each patch the approximate solution is
represented as a linear combination of basis polynomials. Given a
general function, there are many ways to approximate it by
polynomials, and the method we choose is to have the function and its
approximation agree in a discrete set of points, in some way
``sampling'' the approximated function. Those points are called the
\emph{collocation points}. In each patch, the solution is represented
as
\begin{align}
  f(x) \approx \sum_{i=0}^N f_i T_i(x),
  \quad T_i(x) \text{ is a polynomial of degree } i \, ,
\end{align}
therefore, the derivative is approximated by
\begin{align}
  f'(x) \approx \frac {d}{dx} \sum_{i=0}^N f_i T_i(x)
  = \sum_{i,j}^N D_i{}^j f_j T_i(x) \, ,
\end{align}
where~$D_i{}^j$ represents the components of a dense~$N+1\times N+1$
matrix. From this we can infer two important properties of the PSC
method:
\begin{enumerate}
\item For a given number of basis polynomials, the derivative looks
  formally like an~$N$-th order finite difference. Heuristically,
  every time we increase~$N$ we not only make the grid spacing
  smaller, but increase the order of convergence, and therefore we may
  expect exponential convergence in spatial derivatives.
\item The derivative at a single point on the domain depends on the
  value on all other points in the patch. As a result, changing the
  boundary points by hand may have unintended effects everywhere else
  in the patch within one time-step. This leads us to consider an
  alternative way of implementing boundary conditions, which we discuss
  next.
\end{enumerate}
Regarding the question of boundary conditions, we impose Dirichlet
conditions on the incoming physical degrees of freedom (a special case
of maximally dissipative conditions) and constraint preserving
Robin-like conditions at the outer boundary. Besides this, since we
split the domain into spectral elements, we impose Dirichlet
conditions on all incoming variables at these patch boundaries to
match the solutions there. The patching conditions are imposed through
the penalty method~\cite{Hes00}, whereas the Robin (and Von
Neumann if we had any) are imposed through the Bjørhus
method~\cite{doi:10.1137/0916035}. Both work in a similar way: instead
of modifying field values at the boundaries, the evolution equations
for the incoming characteristic variables are modified. This is a weaker
imposition of boundary conditions, but it is informed by studying the
evolution of energy estimates. In simplified and tractable cases,
it has been proven that it leads to a discrete analog of well-posedness
of the initial boundary value problem.

The penalty method consists of modifying the right-hand side of the
evolution of a given characteristic variable $u$ in the boundary as
follows
\begin{align}
  \partial_t u \boundeq (\partial_t u)_{\mathring\Omega}
  +  p\lambda (u - u_{BC})\Theta(-\lambda) \, .
  \label{eq:penalty}
\end{align}
$\Omega$ is the domain, $\mathring\Omega$ its the interior and an
index~$\mathring\Omega$ denotes the use of the same expression used in
the interior. Here~$\lambda$ is the speed of the characteristic, and
the factor of~$\Theta(-\lambda)$ (the Heaviside step function) ensures
we only add modifications to the incoming characteristic
variables. Finally~$p$ is a free parameter fixed by energy
conservation arguments on the numerical approximation, and~$u_{BC}$ is
the value that we want to set the characteristic to at the
boundary. In patch boundaries, $u_{BC}$ corresponds to the value in
the neighboring patch. As discussed in
section~\ref{section:boundary_conditions}, in the outer boundary we
want to set the incoming Newman-Penrose scalar for the electromagnetic
field to be zero. Equivalently, in terms of the characteristic
variables we want to set
\begin{align}
    u_-^A \boundeq 0\,,
\end{align}
which we achieve by the following implementation of the penalty method
\begin{align}
  \partial_t u_-^A \boundeq (\partial_t u_-^A)_{\mathring\Omega}
  -  p(\beta^s + \alpha) u_-^A \, .
\end{align}

We turn our attention now to constraint preserving boundary
conditions. Following~\cite{Lindblom:2005qh}, we can set the desired
Robin-type boundary conditions on the characteristic
variables~$u^{\text{long}, E}_-$, $u^{\text{long}, B}_-$ since
\begin{align}
    \label{eq:constraint_robin_boundary_condition}
    c_-^E \simeq \partial_s u_-^{\text{long}, E}\, ,\quad 
    c_-^B \simeq \partial_s u_-^{\text{long}, B} \, .
\end{align}
The Bjørhus method consists of modifying the evolution equation of a
characteristic variable in the following way
\begin{align}
  \partial_t u \boundeq (\partial_t u)_{\mathring\Omega}
  + \lambda\left(\pd_s u - \pd_s u_{BC}\right)\Theta(-\lambda) \, .
\end{align}
Notice the differences with equation~\eqref{eq:penalty}. Here we use
the longitudinal derivatives and there is no $p$ factor.

In our case, the Bjørhus method for the constraints reduces to
\begin{align}
  \partial_t u^{\text{long}, E}_- \boundeq {} & (\partial_t u_-^{\text{long}, E})_{\mathring\Omega}
  - (\beta^s + \alpha) c^E_- \, ,\\
  \partial_t u^{\text{long}, B}_- \boundeq {} & (\partial_t u_-^{\text{long}, B})_{\mathring\Omega}
  - (\beta^s + \alpha) c^B_- \,
\end{align}
In many cases, we want to simulate axisymmetric systems. In those
cases, we can use the cartoon method~\cite{Alcubierre:1999ab} as
  implemented in~\cite{Pretorius:2004jg}, which allows us to reduce
the simulation domain to a single $\varphi = $ constant slice. The
idea is that, in axisymmetry, all tensors in the problem have the
following symmetry
\begin{align}
    \mathcal L_{\partial_\varphi} T = 0 \, .
\end{align}
Expressing that condition on a coordinate basis, we obtain formulae
for the derivatives perpendicular to the domain in terms of the
derivatives inside the domain. As an example, when~$T$ is a vector,
and considering that we take the slice~$y = 0$ in Cartesian
coordinates, the formula above results in the following condition for
the~$x$ component
\begin{align}
  x\partial_y v^x - y \partial_x v^x + v^y = 0
  \Rightarrow \partial_y v^x = - \frac 1 x v^y \, .
\end{align}

The last point to discuss about the implementation is the initial data
solver. We use the hyperbolic relaxation method
of~\cite{Ruter:2017iph}, which is analogous to traditional parabolic
relaxation solvers, but it is more readily implemented in an evolution
code like \textsc{bamps}.

\section{Numerical Results}
\label{section:numerical_results}

In the experiments presented here, we aim to showcase both the
correctness of the numerical implementation by demonstrating
convergence, as well as the code's ability to reproduce the correct
physics by comparing its results with well understood systems. For
that purpose we perform the following tests:
\begin{itemize}
\item Test convergence to an analytical solution of electrodynamics on a
  flat background.
\item Verify that Reissner-Nordström black hole initial data lead to a
  static simulation.
\item Compute quasinormal mode frequencies of an electromagnetically
  excited black hole that relaxes to Kerr.
\item Run a simulation representative of a critical collapse bisection
  study.
\end{itemize}
Some relevant parameters for each simulation are given in
table~\ref{tab:parameters}. The computational domain is represented in
Fig.~\ref{fig:multipatch}. The approximate costs in core hours of the
different tests are
\begin{itemize}
  \item Eight core hours in average per simulation in the flat
  electrodynamics tests.
  \item 750 core hours for the static Reissner-Nordström test.
  \item 7500 core hours for the relaxation of an electromagnetically
  excited black hole test.
  \item 250 core hours for the critical-collapse-like simulation.
\end{itemize}

\begin{table*}
    \centering
    \setlength{\tabcolsep}{12pt}
    \begin{tabular}{l c c c c c}
    \toprule
     & Flat dipole & $\kappa$ comparison  & Dynamical dipole\\
    \midrule
    Angular resolution & 6 & 8&  8\\
    Transition region resolution & 6 & 6 &  12\\
    Outer shell resolution & 6 & 4 &  10 \\
    Number of collocation points & 11-21 & 15 & 15\\
    AMR & off & off &   on \\
    \addlinespace
    Inner region size$\times\sigma^{-2} $& $2\times2$ & $2\times2$ & $10\times10$\\
    Transition region radius$\times\sigma^{-1}$ & 10 & 10 & 10 \\
    Outer shell radius$\times\sigma^{-1}$ & 20 & 20  & 20\\
    \addlinespace
    Constraint transport & off & on &  on \\
    Damping parameter ($\sigma \kappa$) & 0 & 0-2.5&  1 \\
    \midrule
    \end{tabular}

    \begin{tabular}{l c c c c c}
     &  Reissner-Nordström static & Black hole relaxation\\
    \midrule
    Angular resolution & 6& 8\\
    Outer shell resolution & 100 & 500\\
    Number of collocation points & 11-21 & 15\\
    AMR & off & off \\
    \addlinespace
    Outer shell radius$\times M^{-1}$  & $1.7$ to $201.7$ & $1.9$ to $201.9$\\
    \addlinespace
    Constraint transport & on & on \\
    Damping parameter ($M\kappa$) &  1 & 1\\
    \bottomrule
    \end{tabular}
    \caption{\label{tab:parameters}  Numerical parameters of the
      simulations in the test. There are no data on the inner and
      transition regions of the static Reissner-Nordström and black
      hole relaxation tests since we use excision, which does not have
      those regions (see Fig.~\ref{fig:multipatch})}
\end{table*}
\subsection{Flat background electrodynamics}
\label{subsec:results_flat}
\begin{figure*}
  \centering \includegraphics[width=0.99\linewidth]{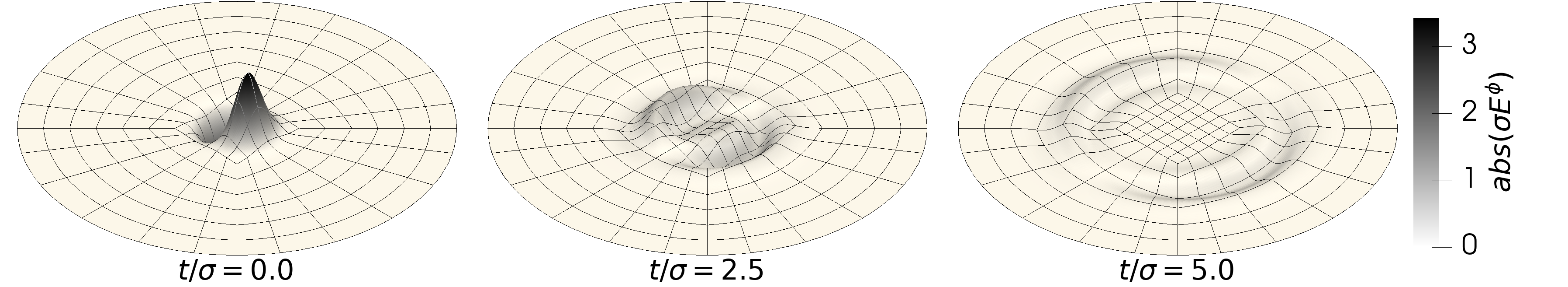}
  \caption{\label{fig:dipole_snapshots}
    Snapshots of~$\sigma E^\phi$
    in the~$xz$ plane at different times. We can see the expected
    behavior of propagation of a dipole wave. This is only a subset of
    the simulation domain; the true boundary is placed at double the
    radius of the plot.}
\end{figure*}
\begin{figure*}
    \centering
    \subfigure{\includegraphics[width=0.32\linewidth]{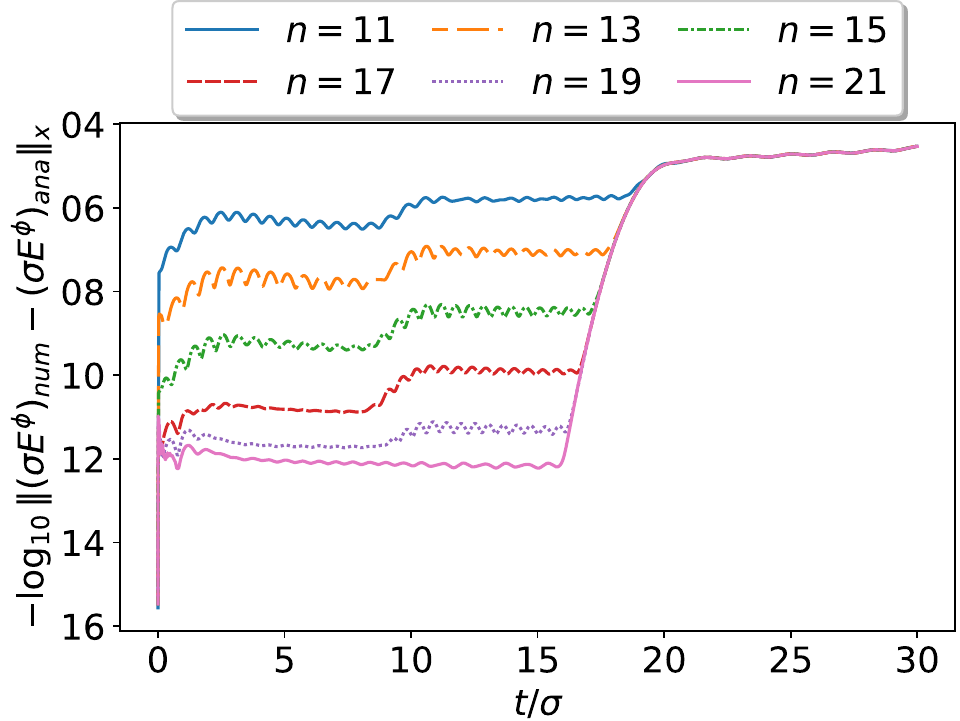}}
    \subfigure{\includegraphics[width=0.32\linewidth]{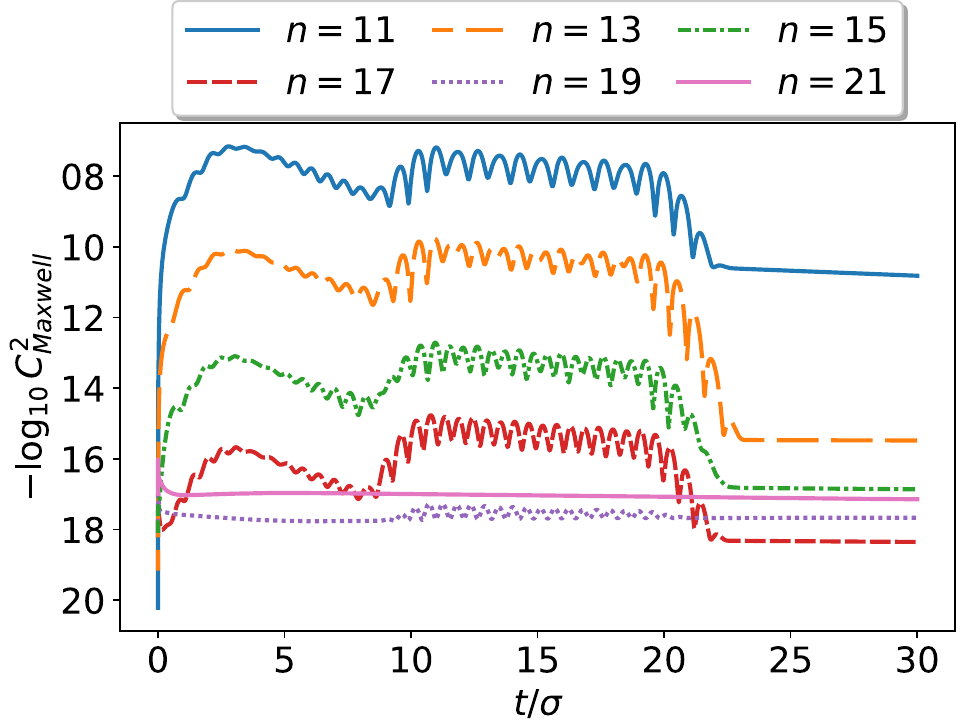}}
    \subfigure{\includegraphics[width=0.32\linewidth]{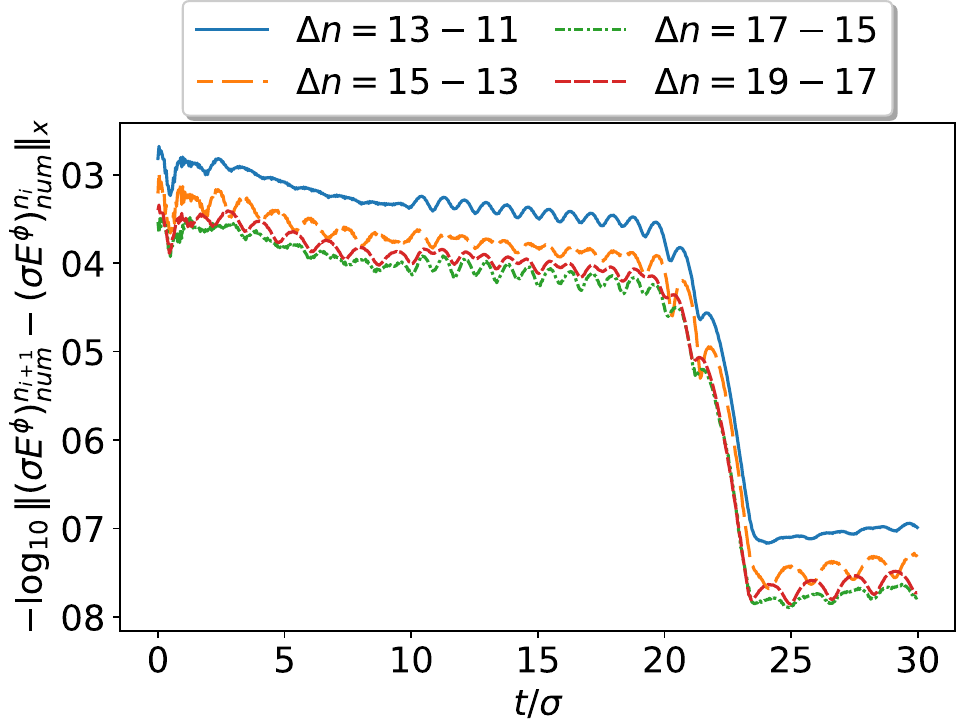}}
    \caption{\label{fig:dipole_results}Convergence of various error
      norms as function of time~$t$ with the number of collocation
      points~$n$ for a simulation of an electromagnetic dipole wave on
      a flat background. Left: Norm of the difference to the analytic
      solution to the Cauchy problem. Observe that here it
      is expected, and observed, that once the outer boundary has a
      significant effect on the numerical solution we can no longer
      see convergence to a solution of the Cauchy problem. Middle:
      Maxwell constraint norm. We observe here that the
      constraints continue to converge even beyond~$t/\sigma\simeq20$, in
      line with our interpretation of the left panel. Right: Norm
      of the difference between subsequent resolutions. The norms of
      the differences are calculated in the~$x$ axis, as in
      equation~\eqref{eq:norm_x_axis}, whereas the constraint monitor
      is calculated with a norm over the whole domain as defined in
      Eq.~\eqref{eq:maxwell_constraint_monitor}.  }
\end{figure*}

This test has the advantages that it only involves the newly
implemented code, and we can compare the simulation results
to analytical solutions. We
choose dipole initial data with a Gaussian profile, which has previously
been used by \cite{Mendoza:2021nwq},
\begin{align}
  \label{eq:electromagnetic_dipole_E}
  E^i\partial_i  &= 8 \mathcal A e^{-(r/\sigma)^2}
  (-y \partial_x + x \partial_y)  \, , \\
  \label{eq:electromagnetic_dipole_B}
  B^i &= 0 \, ,  
\end{align}
where~$\mathcal A$ is the amplitude of the Gaussian profile
and~$\sigma$ is its width. We choose~$\sigma$ to be the characteristic
length scale of the system, and all values presented are in units
of~$\sigma$. The analytical solution that we compare to is
\begin{align}
  E^i\partial_i &= A f_3(u, v)\frac{\partial_\phi}{\sigma}\, ,\\
  B^i\partial_i &= A\frac{z}{r}\left[6f_1(u, v) - f_2(u, v)\right] 
  \partial_r
  + A \left[f_2(u, v) - 2 f_1(u, v)\right] \partial_z\, ,
\end{align}
with the following auxiliary variables
\begin{align}
  u &:= \frac{r-t}{\sigma}\, , \quad v := \frac{r+t}{\sigma}\, ,\\
  f_1(u, v) &:= \frac{u e^{-u^2} - v e^{-v^2}}{(r/\sigma)^2} + \frac{1}{2}
    \frac{e^{-u^2} - e^{-v^2}}{(r/\sigma)^3},\\
  f_2(u, v) &:= 2 \frac{e^{-u^2} - e^{-v^2} - 2 u^2 e^{-u^2} + 2v^2 e^{-v^2}}{r/\sigma},\\
  f_3(u, v) &:= 2 \left(\frac{t_0}{\sigma} \frac{-e^{-u^2} +
  e^{-v^2}}{(r/\sigma)^3} + 2 \frac{u^2 e^{-u^2} + v^2 e^{-v^2}}{(r/\sigma)^2}\right),\\
\end{align}

We represent the solution by the scalar
\begin{align}
    E^\phi := E^\mu (\partial_\phi)_\mu \, ,
\end{align}
where~$\partial_\phi$ is the angular Killing vector. Defining it this
way allows us to compare with the more general case presented in
section~\ref{section:nr_dipole}. Fig.~\ref{fig:dipole_snapshots}
shows snapshots of~$E^\phi$ in this simulation. To test convergence, we
fix the number and layout of the patches, but vary the number of
collocation points. For this test, we do not use constraint damping.
The relevant parameters are presented in table~\ref{tab:parameters}.
Figure~\ref{fig:dipole_results} shows the convergence to the analytical
solution~$E^\phi_\mathrm{ana}$, the convergence of the constraint
norm~$C_{\text{Maxwell}}^2$ (see
Eq.~\eqref{eq:maxwell_constraint_monitor}), and the self convergence of
the numerical approximate solution~$E^\phi_\mathrm{num}$ with respect
to the number of collocation points. The norms of differences of two
functions are calculated in post-processing along the~$x$-axis as
\begin{align}
  \norm{f-g}_x = \sqrt{\int_{x \text{ axis}}
  \left[f(x, 0, 0) - g(x, 0, 0)\right]^2 dx}\,,
  \label{eq:norm_x_axis}
\end{align}
whereas the norms of the constraints are calculated during the
simulation over the entire domain.

The results are consistent with exponential convergence up to the
resolution floor. At high enough resolution, we might expect the
error to be dominated by the polynomial convergence of the RK4 time
integrator, however our results indicate that behavior does not manifest
before the resolution floor is reached.
It is worth noting that, once the pulse reaches the
outer boundary, located at $r = 20\sigma$, the simulation stops converging to the analytical
solution. This is to be expected, since our boundary condition of no
incoming radiation only approximates the solution of the initial value
problem. In the self-convergence test, we find that the solution still
converges after that time is reached.

\begin{figure}
  \centering
  \includegraphics[width=0.5\linewidth]{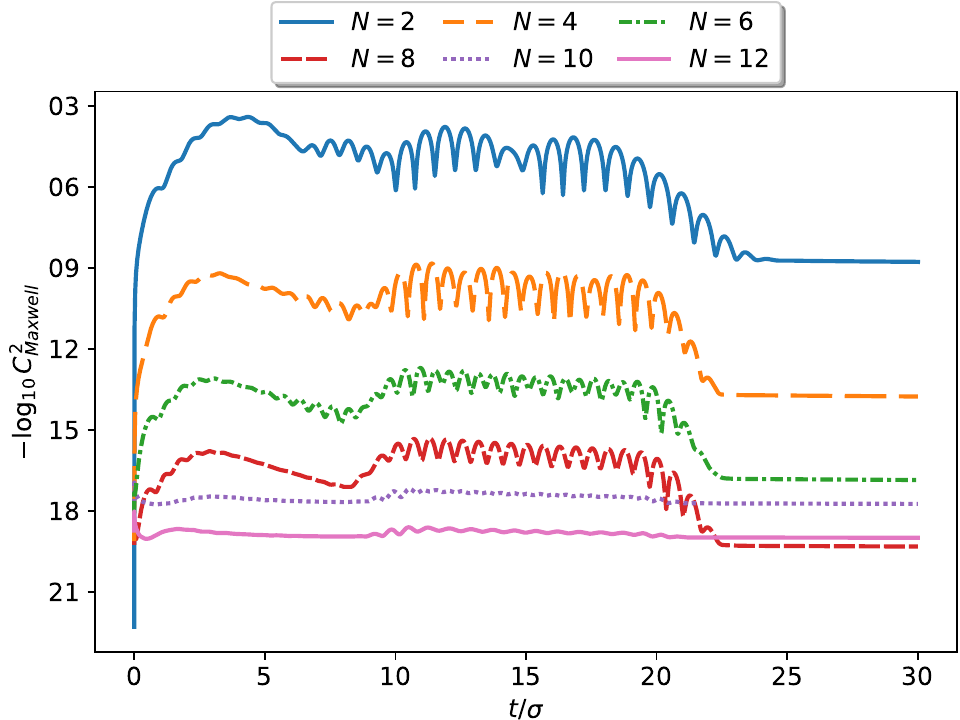}
  \caption{Comparison of the constraint monitor for different number
    of patches, scaled by $N^2$ in all sub-domains. The result is
    consistent with high-order polynomial convergence, and the most
    accurate run shows less error than the ones in
    Fig.~\ref{fig:dipole_results}.
 }
  \label{fig:polynomial_convergence}
\end{figure}

\begin{figure}
  \centering
  \includegraphics[width=0.99\linewidth]{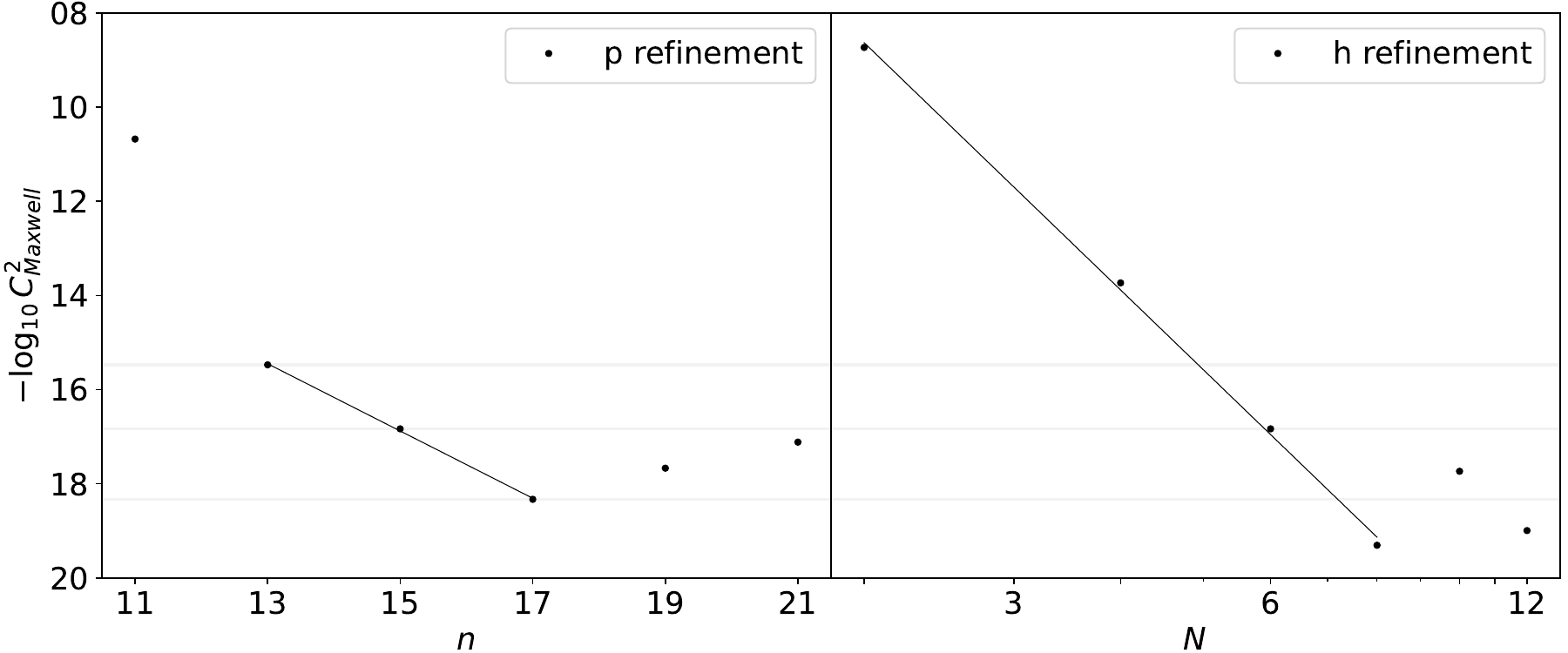}
  \caption{Constraint violation at time $t/\sigma = 25$, for both convergence tests,
  increasing
  the number of collocation points (p refinement) and increasing the number of
  patches (h refinement). Notice that the bottom axis scale is linear in the number of
  collocation points $n$ (left) and logarithmic in the number of patches $N$ (right).}
  \label{fig:hp_convergence}
\end{figure}

Furthermore, we observe that the quantities reach a resolution
floor. This can be seen in the~$n=21$ line of the convergence of the
Maxwell constraint monitor defined in
Eq.~\eqref{eq:maxwell_constraint_monitor}, or in the~$\Delta n = 19-17$
line of the self-convergence plot. The exponential convergence of the
PSC method, along with its higher arithmetic error associated with the
dense matrix multiplications, means that reaching the floor of
resolution is common. Nevertheless, one of the strengths of the approach is that
the error could be reduced even further by increasing the number of
spectral elements. In Fig.~\ref{fig:polynomial_convergence}, we show a
complementary test in which the number of patches is scaled up
instead of the number of collocation points. The result is compatible
with high order polynomial convergence (we are using 15 collocation
points, so we expect the error to decrease with a factor of resolution
to the fifteenth power, up until the time integrator error starts to
dominate). Notably, in this figure we can see that the most precise
run has smaller error than the one presented in
Fig.~\ref{fig:dipole_results}, showing that we can decrease the error
by appropriately choosing both the number of collocation points and
the layout and number of patches. This can also be seen in 
Fig.~\ref{fig:hp_convergence}, which shows the constraint violation at a late
time ($t = 25\sigma$) as a function of resolution for both test cases.
That figure also shows fits to the expected exponential and polynomial
convergence, respectively.

We can also change the constraint damping parameter and see how the
error is effected. In this case the error is monitored as the sum of
the~$L^2$-norms of all the constraints over the whole domain
\begin{align}
\label{eq:maxwell_constraint_monitor}
\begin{split}
 C_{\text{Maxwell}}^2 :={}&  
 \lambda^{-1}\left(\norm{Z_E}^2 + \norm{Z_B}^2\right)
                   +   \lambda\left(\norm{\calg_E}^2 + \norm{\cal G_B}^2\right)
\end{split}                   
                   \nonumber\\
                   ={}& \int d^3x\; \sqrt{\gamma}\left(\lambda^{-1} Z_E^2 + \lambda^{-1}Z_B^2
                  + \lambda \calg_E^2 +  \lambda\mathcal{G}_B^2\right) \, .
\end{align}
Where~$\lambda$ is the characteristic length scale of the system
($\sigma$ in this case and $M$ in the black hole simulations). The
powers are chosen so that the constraint monitor is dimensionless.

\begin{figure}
  \centering
  \includegraphics[width=0.5\linewidth]{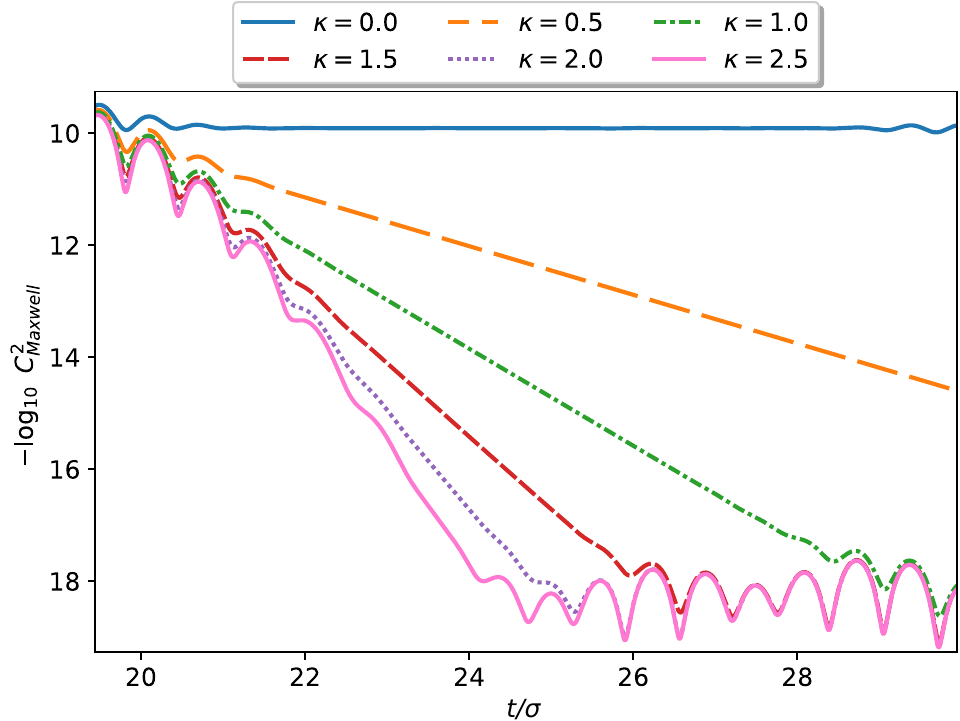}
  \caption{Constraint norm as a function of
    time $t$ in a simulation of an electromagnetic dipole wave on a flat background for
    different choices of the constraint damping parameter $\kappa$.
 }
  \label{fig:dipole_kappa}
\end{figure}

Figure~\ref{fig:dipole_kappa} shows the evolution of the constraint norm
for different values of the constraint damping parameter. We can see that when the
constraints are exponentially damped, the decay rate increases
with~$\kappa$.

These results give us good confidence that the flat electrodynamics
part of the code is well implemented. The terms related to curved
spacetime are evaluated in the following section.

\subsection{Static Reissner-Nordstr\"om simulation}

\begin{figure*}
    \centering
    \subfigure{\includegraphics[width=0.32\linewidth]{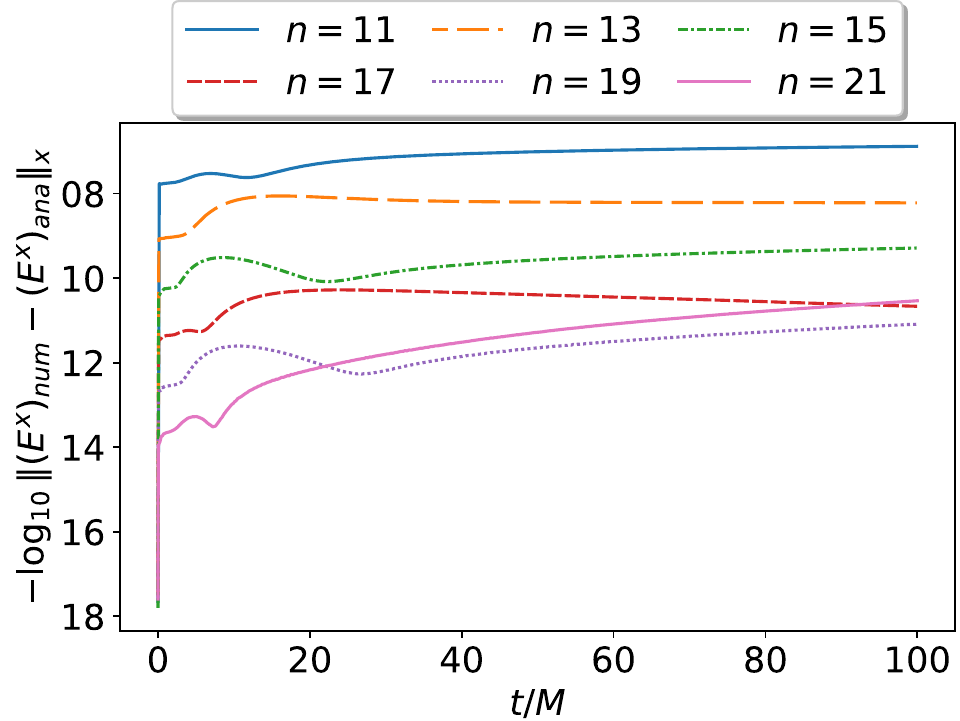}}
    \subfigure{\includegraphics[width=0.32\linewidth]{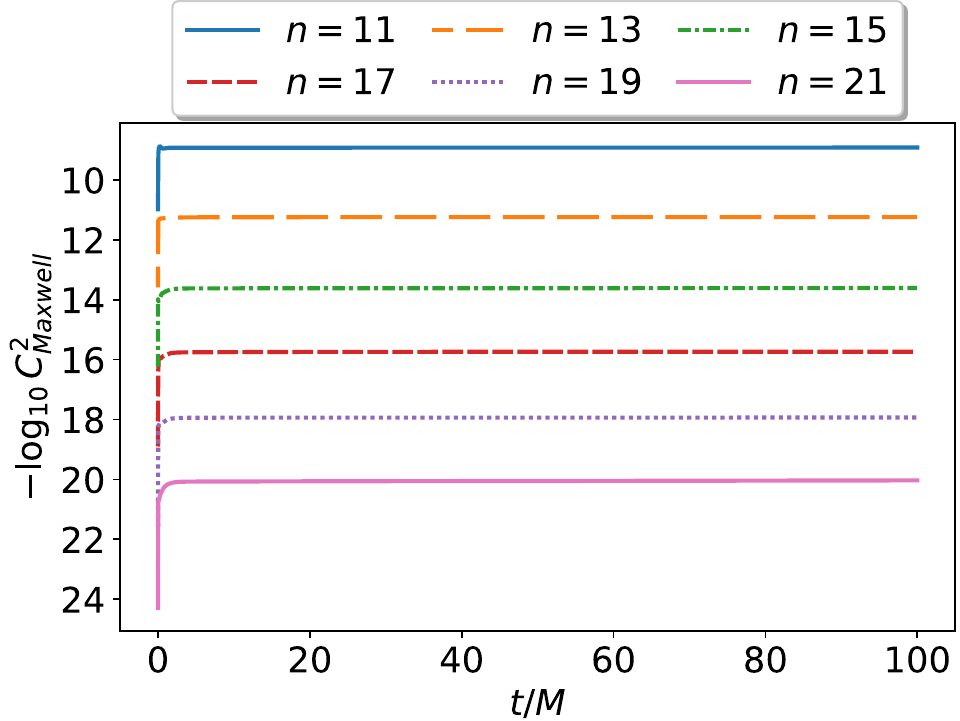}}
    \subfigure{\includegraphics[width=0.32\linewidth]{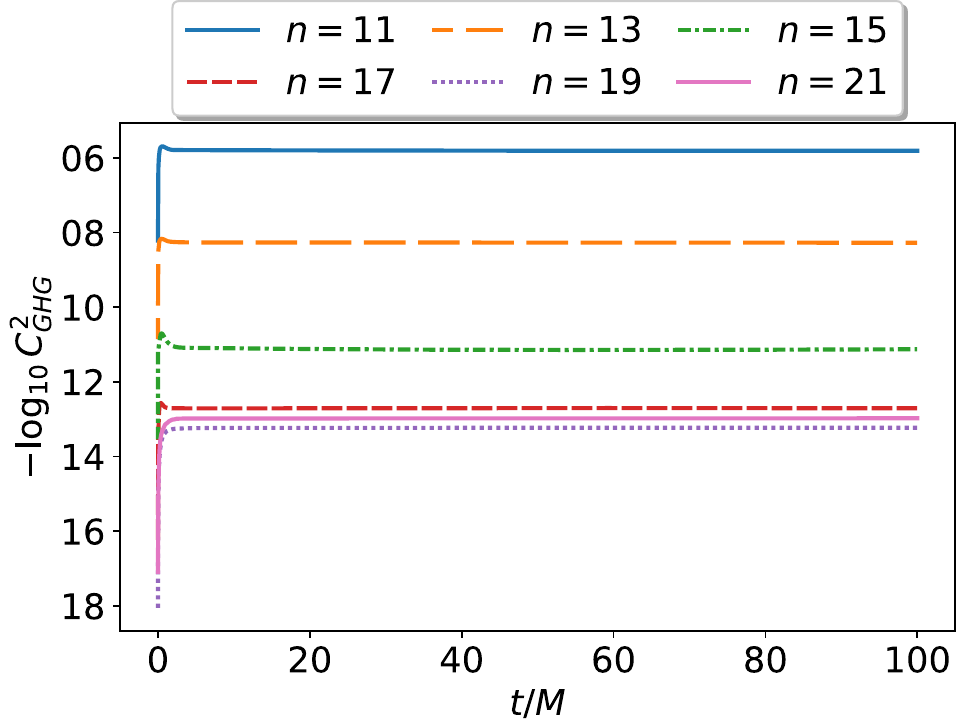}}
    \caption{Convergence of various error norms as function of
      time~$t$ with the number of collocation points~$n$ for a
      simulation of a Reissner-Nordström black hole.  Some relevant
      parameters of the simulation are given in
      table~\ref{tab:parameters}. Left: Norm of the difference to the
      analytic solution taken over the~$x$ axis only. Middle: Maxwell
      constraint norm. Right: Norm of the GHG constraint monitor.}
    \label{fig:static-RN}
\end{figure*}

In order to test the interaction of gravity and electromagnetism in
the simplest case, we set up initial data corresponding to a
Reissner-Nordström black hole and check that we have both a convergent
and static evolution.

We set up initial data in Kerr-Schild coordinates~\cite{Kerr:1965vyg}
because these are smoothly horizon penetrating, and we set an excision
surface at~$r=1.7M$. The initial data for the electromagnetic fields
are
\begin{align}
  E^i = \frac 1 {\sqrt{1+H}} \frac{Q}{r^2} (\partial_r)^i , \quad B^i =0 \, .
\end{align}
where
\begin{align}
    H = \frac {2M}{r} - \frac{Q^2}{r^2} \, ,
\end{align}
is the free function that appears in Kerr-Schild coordinates. The
initial data for the gravitational field are given by
\begin{align}
    ds^2 ={}& - (1-H) d\tau^2 + (1+H) dr^2 +  H(drd\tau + d\tau d r)
     +  r^2d\Omega^2 \,, \\ 
     K_{ij} ={}& \frac 12 \frac{H+2}{\sqrt{H+1}} \partial_r H {(dr^2)}_{ij} \, .
\end{align}
As stated above, we employ a first order reduction of the GHG
  formulation of GR. In this experiment, we chose gauge source
functions that correspond to the lapse and shift of a
Reissner-Nordström spacetime in Kerr-Schild coordinates, which have
zero time derivative in the initial data.

For the convergence test we again keep the number of patches fixed and
vary the number of collocation points.  Figure~\ref{fig:static-RN} shows
the convergence to the analytical solution $E^\phi_\mathrm{ana}$, the
convergence of the Maxwell system constraint norm
$C_{\text{Maxwell}}^2$, and the convergence of the GHG constraint
monitor with respect to the number of collocation points.  The GHG
constraint monitor was introduced in~\cite{Hilditch:2015aba} and
quantifies both the Hamiltonian and Momentum constraints as well as
the violation of the order reduction and harmonic constraints. The figure
shows that the error is approximately static, and that the solution
converges exponentially with resolution.

\subsection{Relaxation of an electromagnetically excited black hole}

\begin{figure}
    \centering
    \includegraphics[width=0.5\linewidth]{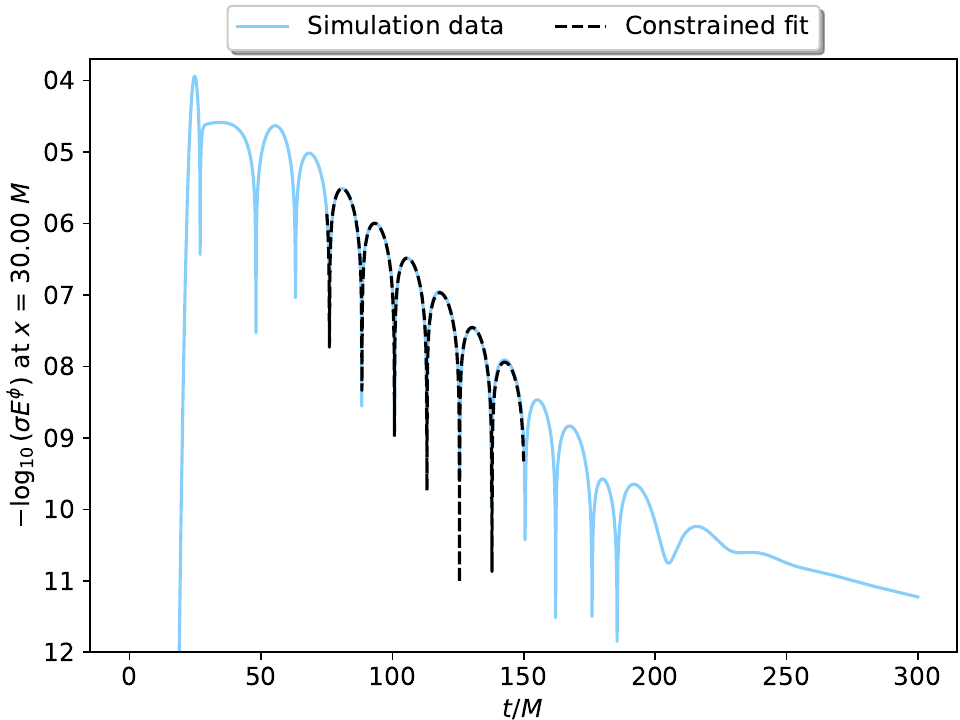}
    \caption{Comparison of the simulation data with the corresponding
      first quasinormal mode in a Kerr black hole. We can see good
      agreement.}
    \label{fig:qnms}
\end{figure}

\begin{figure*}
    \centering
    \subfigure{\includegraphics[width=0.99\linewidth]{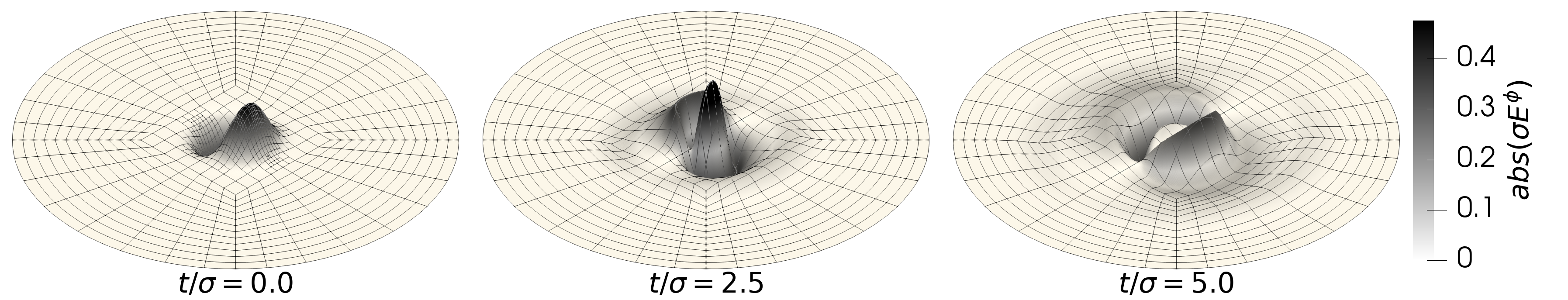}}
    \subfigure{\includegraphics[width=0.99\linewidth]{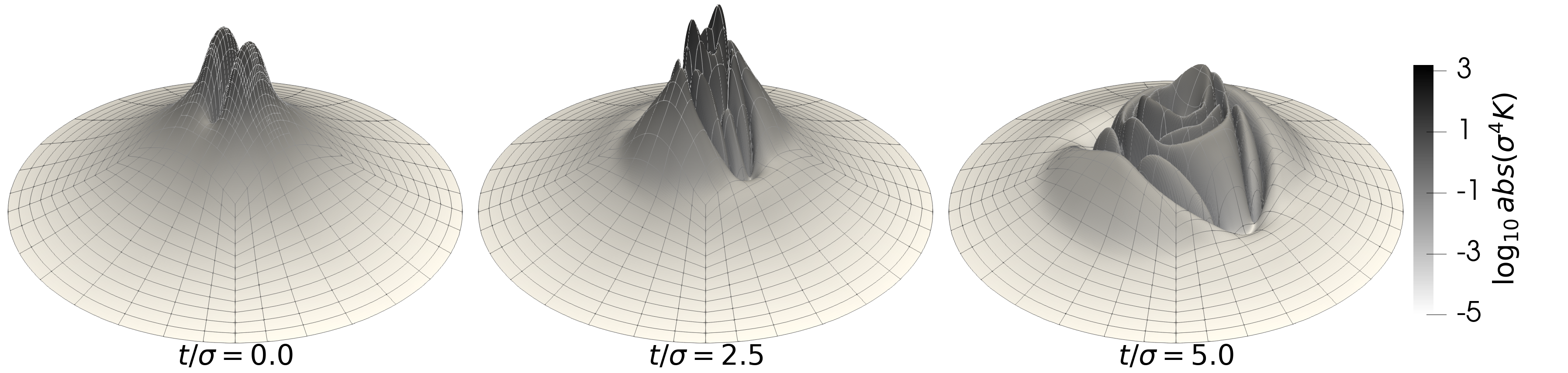}}
    \caption{\label{fig:nr_dipole_snapshots}
      Top panel: Snapshots of
      $\sigma E^\phi$ at different times, in the $xz$ plane for a
      electromagnetic amplitude of $\sigma\mathcal A = 0.7$.\\
      Bottom panel:
      Snapshots of
      the logarithm of the Kretschmann scalar, as given by
      Eq.~\eqref{eq:Kretschmann_log}, at different times, in the $xz$
      plane.}
\end{figure*}

\begin{figure}
    \centering
    \includegraphics[width=0.5\linewidth]{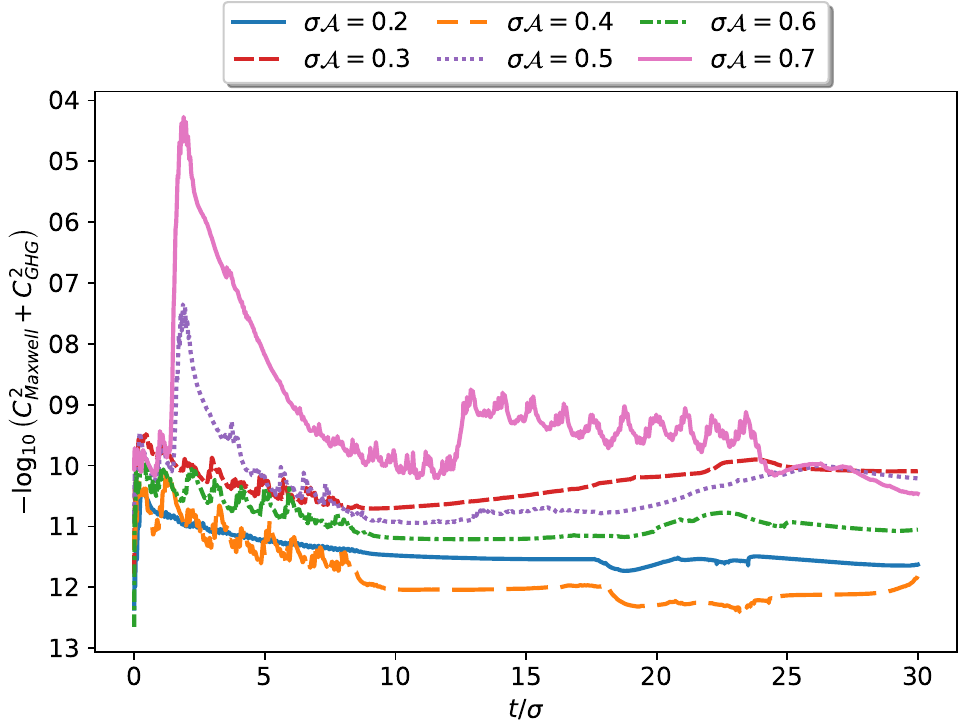}
    \caption{$L^2$-norm of the constraint monitors throughout a fully
      nonlinear evolution, for different amplitudes of the initial data.}
    \label{fig:constraint_comparisons}
\end{figure}

In this test, we set up initial data that is conformally Kerr, with an
electromagnetic excitation, and measure the frequency of the
fundamental mode of relaxation. It is a fully non-linear simulation
in which the geometry is evolved at the same time as the
electromagnetic field.  For the initial data, we use the CTS solver
with freely specifiable variables set to those of a Kerr black
hole. For the electromagnetic variables, we choose
\begin{align}
  \tilde E^i \partial_i = \mathcal{A} e^{-(r-r_0)^2/\sigma^2}
  (-y\partial_x + x \partial_y) \, ,
  \quad \tilde B^i = 0 \, .
\end{align}
These are analytic solutions to the electromagnetic constraints in the
conformal Kerr geometry, since the electric field only points in the
azimuthal direction and the spacetime is axisymmetric and hence
\begin{align}
  \bar D_i \tilde E^i = \partial_\varphi \tilde E^\varphi +
  \frac 12 \tilde E^\varphi \partial_\varphi \log\bar\gamma = 0 \, .
\end{align}
We expect the end-state of evolution to be a Kerr black hole. In
particular, one might wonder if the electromagnetic excitation would
lead to Kerr-Newman instead, however there is no charge in our initial data, and
that must remain so during evolution. After the initial data are set
up, the evolution is performed until $t = 250M$, with the boundaries
at $r=500M$, since we find that this simulation is very sensitive to
boundary effects. The results are presented in Fig.~\ref{fig:qnms},
where we compare them with the analytical expectation of the first
quasi-normal mode around a Kerr black hole. The frequency to compare
is taken from~\cite{Bertipage, Berti:2005ys, Berti:2009kk}.  We find
good agreement of the simulation and the theoretical expectation.

\subsection{Electromagnetic multipoles in a dynamical
  spacetime}
\label{section:nr_dipole}

As a last test, we evolve a strong gravity case with electromagnetic
multipole initial data.  In this evolution, the electromagnetic field
interacts with itself through gravity and ends up dispersing. This is
a good representation of the type of simulations that are required to
study critical collapse, and is a good test for the full numerical
setup, with the interaction with AMR being of particular interest.

For the initial data, we set the conformal electromagnetic fields to
be the same as in Eqs.~\eqref{eq:electromagnetic_dipole_E}
and~\eqref{eq:electromagnetic_dipole_B}, with an amplitude to width
relation of $\sigma\mathcal A = 0.7$, and solve the gravity
constraints with the choices for the freely specifiable variables
corresponding to flat spacetime. This amplitude is not enough to
collapse, and is not well tuned to the threshold of collapse, but
nevertheless allows us to see comparable strong-field dynamics before
dispersing.

The top panel of Fig.~\ref{fig:nr_dipole_snapshots} shows snapshots of~$E^\phi$ in
this simulation. Notice the difference to the evolution on a flat
background that is shown in Fig.~\ref{fig:dipole_snapshots}. For instance,
the electromagnetic field takes longer to disperse,
which we can interpret as the effect of self-gravitation that keeps
the wave packet together for longer. Based on the knowledge gathered
in other matter models~\cite{Gundlach:2007gc}, we expect this time to
increase as we approach the threshold of collapse.
The bottom panel of Fig.~\ref{fig:nr_dipole_snapshots} shows snapshots of
the Kretschmann scalar for the same evolution. In particular, we look
at its logarithm in units of $\sigma$
\begin{align}
\label{eq:Kretschmann_log}
    \log(\sigma^4 \left| R_{\mu\nu\rho\sigma} R^{\mu\nu\rho\sigma}\right|) \, .
\end{align}
Even with this poor degree of tuning to the threshold of collapse, we
see complicated dynamics in the curvature, which reaches large values
before dispersing.

Finally, in Fig.~\ref{fig:constraint_comparisons} we show the evolution
of the constraint monitors. These integrals are several orders of
magnitude below the Kretschmann scalar. Consequently, they are small
enough and we can trust the simulation results. These measures of
error will be more challenging to keep under control with more tuning,
but past
experience~\cite{Fernandez:2022hyx,Cors:2023ncc,Marouda:2024epb}
indicates that the combination of constraint damping with the
\textsc{bamps} AMR scheme will go a long way to do so efficiently.

\section{Conclusions}
\label{section:conclusions}

Motivated by future investigation into various aspects of the
electrovacuum dynamics, we have presented a formulation of the
Einstein-Maxwell system as an IBVP. We showed that it has several
desirable properties for simulations, including symmetric
hyperbolicity and boundary stability. Our main goal in choosing both a
mathematical formulation and a discretization scheme was to maximize
the accuracy to cost ratio.

To the best of our knowledge, we have developed the first PSC
implementation of the Einstein-Maxwell system, and this will allow us
to obtain accurate results in extreme scenarios. We have presented a
suite of tests, demonstrating the correctness of the different parts
of the code. An evolution of a dipole electromagnetic wave on a fixed
flat background verified the flat terms on the new evolution code. A
static Reissner-Nordström simulation verified the interaction with
gravity in a simple setting. A measurement of electromagnetic
quasinormal modes around a Kerr black hole agreed with the existing
literature, thereby validating the physical results of our
code. Finally, a strong gravity but subcritical electromagnetic dipole
run verified the interaction of the new code with the extensive
numerical infrastructure, e.\,g.\@ adaptive mesh-refinement in a
highly dynamical setting. Our conclusion
is that the code is working as expected,
and is able to evolve the systems of interest.

In terms of the future of the code, an interesting avenue would be to
include interaction with a scalar field in order to expand the systems
that we can study as well as to provide an interface with recent
mathematical work (for instance that of~\cite{Kehle:2024vyt}).

The code and the tests here presented provide a good basis to study
topics such as critical collapse, but also the dynamics of the
Einstein-Maxwell system in more generality, with particular examples
being the mode-mixing of electromagnetic waves and gravitational waves
in a strong background magnetic field (the Gertsenshtein
effect~\cite{Gertsenshtein}), or the electromagnetic contributions to
quasinormal mode ringing.

\section*{Acknowledgments}

We are grateful to Florian Atteneder and Fernando Abalos for helpful
discussions and to Stephanie Thabuis for revising the text. 

\paragraph{Funding information}
J.\,E.\,P. and H.\,R.\,R. acknowledge financial support provided
under the European Union’s H2020 ERC Advanced Grant ``Black holes:
gravitational engines of discovery'' grant agreement no.\@
Gravitas–101052587.  Views and opinions expressed are, however, those of
the authors only and do not necessarily reflect those of the European
Union or of the European Research Council.  Neither the European Union
nor the granting authority can be held responsible for them.  
D.\,H. was
partially supported by the FCT (Portugal) 2023.12549.PEX grant. The
authors were partially supported by FCT (Portugal) projects
UIDB/00099/2020 and UIDP/00099/2020.
The authors thankfully acknowledge
the computer resources, technical expertise and assistance provided by
CENTRA/IST. Computations were performed at the cluster
``Baltasar-Sete-Sóis'' and supported by the H2020 ERC Advanced Grant
``Black holes: gravitational engines of discovery'' grant agreement
no.\@ Gravitas–101052587.
The authors gratefully acknowledge the Gauss Centre for Supercomputing
e.V. (www.gauss-centre.eu) for funding this project by providing
computing time on the GCS Supercomputer SuperMUC-NG at Leibniz
Supercomputing Centre (www.lrz.de).

\bibliographystyle{SciPost_bibstyle}
\bibliography{main.bbl}

\end{document}